\documentclass[useAMS,usenatbib]{mn2e}

\usepackage{graphicx}
\usepackage{amssymb}
\usepackage{amsmath}
\usepackage{ulem}

\usepackage{ulem}
\usepackage[usenames,dvipsnames]{color}

\newcommand\msun{\, \rm M_\odot}
\newcommand\pc{{\, \rm pc}}

\newcommand\myr{{\, \rm Myr}}

\def\gtsima{$\; \buildrel > \over \sim \;$}
\def\ltsima{$\; \buildrel < \over \sim \;$}
\def\gtrsim{\lower.5ex\hbox{\gtsima}}
\def\lesssim{\lower.5ex\hbox{\ltsima}}

\def\apj{ApJ}%
\def\mnras{MNRAS}%
\title[BH-BH binaries in YSCs]{Dynamics of stellar black holes in young star clusters 
with different metallicities - II. Black hole-black hole binaries}
\author[Ziosi et al.]  
  {Brunetto Marco Ziosi$^{1,}{}^{2}$\thanks{E-mail: brunetto.ziosi@gmail.com}, 
  Michela Mapelli$^{2}$, 
  Marica Branchesi$^{3,}{}^{4,}{}^{5}$,
  Giuseppe Tormen$^{1}$\\
  $^{1}$Universit\`a degli Studi di Padova, Vicolo dell'Osservatorio 3, I--35122, Padova, Italy\\
  $^{2}$INAF-Osservatorio Astronomico di Padova, Vicolo dell'Osservatorio 5, I--35122, Padova, Italy\\
  $^{3}$Universit\`a degli Studi di Urbino ``Carlo Bo'', Dipartimento di Scienze di 
  Base e Fondamenti, Piazza della Repubblica 13, I--61029, Urbino, Italy \\
  $^{4}$INAF-Osservatorio Astronomico di Roma, Via Frascati 33, I--00040, Monte Porzio Catone (RM),  Italy\\
  $^{5}$INFN, Sezione di Firenze, I-50019 Sesto Fiorentino, Firenze, Italy\\
  }

\begin{document}

\date{}

\pagerange{\pageref{firstpage}--\pageref{lastpage}} \pubyear{2002}

\maketitle

\label{firstpage}

\begin{abstract}
In this paper, we study the formation and dynamical evolution of black hole-black hole  (BH-BH) binaries 
in young  star clusters (YSCs), by means of N-body simulations. 
The simulations include metallicity-dependent recipes for stellar evolution and stellar winds, 
and have been run for three different metallicities ($Z = 0.01, 0.1\text{ and 1 Z}_\odot$). 
Following recent theoretical models of wind mass-loss and core-collapse supernovae, 
we assume that the mass of the stellar remnants depends on the metallicity of 
the progenitor stars. 
We find that BH-BH binaries form efficiently because of dynamical exchanges: in our simulations, 
we find about 10 times more BH-BH binaries than double neutron star binaries. 
The simulated BH-BH binaries form earlier in metal-poor YSCs, which  host more massive black 
holes (BHs) than in metal-rich YSCs. 
The simulated BH-BH binaries have very large chirp masses (up to 80 M$_\odot$), 
because the BH mass is assumed to depend on metallicity, 
and because BHs can grow in mass due to the merger with stars. 
The simulated BH-BH binaries span a wide range of orbital periods ($10^{-3}-10^7$ yr), 
and only a small fraction of them (0.3 per cent)
is expected to merge within a Hubble time. We discuss the estimated merger rate 
from our simulations and the implications for Advanced VIRGO and LIGO.
\end{abstract}



\section{Introduction}
\label{sec:intro}
Most stars are expected to form in young star clusters (YSCs, \protect\citealt{carpenter2000}; 
\protect\citealt{ladalada2003}; \protect\citealt{porras2003}). Like globular clusters (GCs), 
the densest YSCs are collisional systems: their two-body relaxation timescale is  
shorter than their lifetime, 
and they undergo intense dynamical evolution. 
On the other hand, YSCs are considerably different from GCs: the former have generally 
lower mass ($<10^{5}$ M$_\odot{}$) and smaller size (half-mass radius $r_{\rm hm}\lesssim{}1$ pc)
 than the latter (see e.g. \citealt{pz2010}, for a recent review). 
This explains why the central relaxation time of YSCs is $\sim{}10-50$ Myr, orders of magnitude 
shorter than that of GCs (e.g. \citealt{pz2004}). YSCs populate the disc of late-type galaxies, 
while GCs are sphe\-ri\-cal\-ly distributed in the host-galaxy halo. Finally, GCs are old ($\gtrsim{}12$ Gyr) 
and long-lived systems, whereas YSCs are young and short lived: most of them dissolve in the disc of the 
host galaxy in $\le{}10^8$ yr (e.g. \citealt{k2011}). 

Thus, the stellar content of dissolved YSCs is expected to build up a considerable fraction of the 
field population of the host galaxy. This must be taken into account when mo\-del\-ling the evolution of 
binary stellar systems in the galactic field: a large fraction of these binaries likely formed in YSCs, 
and then evolved through intense dynamical interactions, before being ejected into the field after the 
disruption of the parent YSC. This scenario is important for the study of stellar black hole (BH) binaries. 
In \citet[][hereafter Paper I]{mapelli2012}, we studied the formation and the dynamical evolution of 
accreting BH binaries in YSCs. We found that dynamical interactions in YSCs have a significant impact 
on the expected population of X-ray sources powered by BHs.

In the current paper, we study the formation and the dynamical evolution of black 
hole-black hole (BH-BH) binaries in YSCs. For the sake of completeness, we will compare 
the evolution of BH-BH binaries with that of neutron star-neutron star (NS-NS) binaries and with 
that of binaries composed of a BH and a neutron star (NS) in YSCs. BH-BH, NS-NS 
and NS-BH binaries are among the most promising sources of gravitational waves 
(GWs) detectable by ground-based detectors (e.g. \citealt{peters1964}; \citealt{abra1994}). 
Understanding the demographics of such double compact object binaries (DCOBs) is 
particularly important in light of the forthcoming second-generation ground-based GW detectors, 
Advanced LIGO and VIRGO  \citep{harry2010, acernese2009, accadia2012}.

The dynamics of YSCs can influence the formation and evolution of BH-BH binaries  
in three different ways:
 (i) dynamical friction causes the BHs (which are more massive than most stars) 
 to sink to the denser YSC core, where they have a higher probability 
 to interact with other BHs (e.g. \citealp{sigurdsson1996});
  (ii) three-body encounters (i.e. close encounters between a binary and a single star) 
  change the binary orbital properties: if the binary is hard (i.e. if its binding energy 
  is higher than the average kinetic energy of a star in the cluster\footnote{A binary can be classified as hard if its
binding energy is higher than the average kinetic energy of stars in the cluster, 
that is 
\begin{equation}
 \frac{G\,{}m_1\,{}m_2}{2\,{}a}\gtrsim \frac{1}{2}\langle m\rangle\sigma^2,
\end{equation}
where $G$ is the gravitational constant, $m_1$ and $m_2$  are the mass of the 
primary member and the mass of the secondary member of the binary, respectively, 
while $\langle m\rangle$ and $\sigma$ are the average mass and velocity 
dispersion of a star in the star cluster.}), three-body  
  encounters tend to shrink the binary semi-major axis \citep{heggie1975};
  (iii) dynamical exchanges (i.e. three-body interactions in which one of 
  the members of the binary is replaced by the single star) enhance the formation of BH-BH binaries. 
  In fact, the probability for a single star with mass $m_3$ to replace a binary member 
  is higher if $m_3\ge{}m_1$ or $m_3\ge{}m_2$ (where $m_1$ and $m_2$ are the masses of 
the former binary members, see \citealt{hills1989} and \citealt{hills1992}). 
As BHs  are more massive than most stars, they 
  efficiently acquire companions through dynamical exchanges.

 Previous studies investigated the formation and evolution of DCOBs 
 either in  GCs, via Monte Carlo codes (e.g. \citealt{oleary2006}; 
 \citealt{sadowski2008}; \citealt{downing2010}; \citealt{downing2011}; \citealt{clausen2012}), 
 or in the field, using  population synthesis simulations of isolated binaries 
 (e.g. \citealt{b2002}; \citealt{voss2003}; \citealt{pfahl2005}; \citealt{dewi2006}; 
 \citealt{b2007}; \citealt{b2010a}; \citealt{dominik2012}). Our study provides a new 
 perspective on this subject: we study the formation of BH-BH binaries in YSCs, by using 
 direct N-body simulations coupled with up-to-date stellar and binary evolution  recipes. 
 The paper is organized as follows. 
In Section \ref{sec:MethodsAndSimulations}, we briefly describe our simulations. In Section \ref{sec:results},
we present our results. Section \ref{sec:Discussion} is devoted to discuss the results and to
compare them with previous work. Our conclusions are presented in Section 
\ref{sec:Conclusions}.


\section[]{Methods and simulations}
\label{sec:MethodsAndSimulations}
 \begin{table}
\begin{center}
 \caption{Summary of initial YSC properties}
   \begin{tabular}{ll}
  \hline
  Parameter & Value \\
  \hline
  $W_0$ & 5\\
  $N_\ast$ & 5500\\
  $r_{\rm c}$ (pc) & 0.4 \\
  $c \equiv \log_{10}(r_{\rm t}/r_{\rm c})$ & 1.03\\
  IMF & Kroupa (2001) \\
  $m_{\rm min}$ (M$_\odot$) & 0.1 \\
  $m_{\rm max}$ (M$_\odot$) & 150 \\
  $Z$ (Z$_{\rm \odot}$) & 0.01, 0.1, 1\\
  $t_{\rm max}$ (Myr) & $100$\\
  $f_{\rm PB}$ & 0.1 
  \vspace{0.1cm}\\
\hline
 \end{tabular}
   \label{tab:SCSummary}
\begin{flushleft}
\footnotesize{$W_0$: central dimensionless potential in the \citet{king1966} model; 
$N_{\ast}$: number of stars per YSC;  $r_{\rm c}$: initial core radius; 
$c\equiv{}\log{}_{10}{(r_{\rm t}/r_{\rm c})}$: concentration 
($r_{\rm t}$ is the initial tidal radius); IMF: initial mass function; 
$m_{\rm min}$ and $m_{\rm max}$: minimum and maximum simulated stellar mass, 
respectively; $Z$: metallicity of the YSC (in our simulations, we assume Z$_\odot{}= 0.019$); $t_{\rm max}$: duration of each simulation 
(in Myr); $f_{\rm PB}$: fraction of PBs,  defined as the number of PBs in each YSC 
divided by the number of `centres of mass' (CMs) in the YSC. In each simulated YSC, 
there are initially 5000 CMs, among which 500 are designated as `binaries' and 
4500 are `single stars' (see \citealt{downing2010} for a description of this formalism). 
Thus, 1000 stars per YSC are initially in binaries.}
\end{flushleft}
\end{center}
 \end{table}

 The simulations analysed in this paper adopt the same technique as described in paper~I.
 In particular, we used a modified version of the {\sc starlab} public software 
 environment (see \citealt{pz2001}).
Our upgraded version of {\sc starlab} includes
(i) 
 analytic formulae for stellar evolution 
as a function of mass and metallicity
 \citep{hurley2000}, (ii) 
metallicity-dependent stellar winds for main sequence
 \citep{vink2001} and evolved stars \citep{vink2006}, 
 and (iii) 
 the possibility that massive BHs form by direct collapse, i.e. with a weak or no supernova (SN) explosion
 (e.g. \citealt{fryer1999}; \citealt{fek2001}; \citealt{mapelli2009}; \citealt{b2010b}; \citealt{fryer2012}).  

According to these recipes, if the final mass of the progenitor star (i.e. the mass before the collapse), 
is $>40$ M$_\odot{}$, we assume that the SN fails and that the star collapses quiet\-ly to a BH. 
 As the final mass of a massive star is higher  
at low metallicity, because of the weaker stellar winds, 
BH masses are allowed to be higher at low metallicity. In particular, the BH mass depends on the metallicity and on 
the zero age main sequence (ZAMS) mass of the progenitor as  described in Fig. 1 of paper~I.  
In this scenario, BHs with mass up to $\sim{}80$  M$_\odot{}$ ($\sim{}40$ M$_\odot{}$) can form 
if the metallicity of the progenitor is $Z \sim{} 0.01\text{ Z}_\odot$ ($Z\sim{} 0.1\text{ Z}_\odot$). 
The ma\-xi\-mum BH mass at  $Z \sim{}\text{ Z}_\odot$ is $23$   M$_\odot{}$. 
This is higher than assumed in previous studies (e.g. \citealt{b2010b}), 
but is still consistent with the observations, given the large uncertainties (e.g. \citealt{ozel2010}).

NSs and BHs that form from a SN explosion receive a natal kick in a random 
direction. The natal kick of NSs is chosen randomly from the distribution 
$P(u)=(4/\pi{})\,{}(1+u^2)^{-2}$, where $u=v/\tilde{v}$, $v$ is the modulus of 
the velocity vector of the NS and $\tilde{v}=600$ km s$^{-1}$ 
\citep{hartman1997, pz2001}. 
The natal kick of BHs is drawn from the same distribution, but is normalized by
a factor $f_{\rm kick}=(m_{\rm NS}/m_{\rm BH})^{1/2}$ (where $m_{\rm BH}$ 
is the BH mass and $m_{\rm NS}=1.3\,{}{\rm M}_\odot{}$ is the typical NS mass). Instead, BHs that form from quiet 
collapse are assumed to receive no natal kick (see \citealt{fryer2012}).

Furthermore, {\sc starlab}  includes recipes for binary evolution, such as mass 
transfer (via wind accretion and via Roche lobe overflow), 
tidal circularization, magnetic braking, and also orbital decay and circularization by GW 
emission (see \citealt{pz1996}; \citealt{pz2001}).

We doubled the simulation sample with respect to paper~I: we have 600 
N-body realizations of YSCs (1/3 of them with solar metallicity, 1/3 with metallicity 
$Z=0.1\text{ Z}_\odot$, and the remaining 1/3 with  $Z=0.01\text{ Z}_\odot$).
Half of the simulations were already presented in paper~I, whereas the remaining are new simulations.

The simulated YSCs are initially modelled with 5000 centres of mass
(single stars or binaries), following a King profile with central dimensionless
potential $W_0=5$. The core density at the beginning of the simulation is 
$\rho_\mathrm{C} \sim 2\times 10^3\msun pc^{-3}$. We chose a primordial binary 
fraction of $f_{\mathrm{PB}} = 0.1$ so the total number of stars is $N_\ast{} = 5500$. 
The total mass of a single YSC is $M_{\mathrm{TOT}}\sim 3-4\times 10^3\msun$. 
The single stars and the primary stars ($m_1$) of the binaries follow a Kroupa initial mass function 
\citep[IMF, ][]{kroupa2001} with minimum and  maximum mass equal to 0.1 and 150$\msun$, respectively. 
The masses of the secondaries ($m_2$) are generated according to a uniform 
distribution between $0.1m_1$ and $m_1$. The initial semi-major axis $a$ of the binaries 
are drawn from a log-uniform distribution $f(a)\propto 1/a$ between $R_\odot$ 
and $10^5R_\odot$, for consistency with the observation of binaries 
in the Solar neighbourhood \citep{kraicheva1978, duquennoy1991}. Values of $a$ 
leading to a periastron
separation smaller than the sum of the radii of the two stars in the binary were 
discarded. We randomly select the initial eccentricity
from a thermal distribution $f(e) = 2\,{}e$ in the range [0, 1] \citep{heggie1975}.

The central relaxation timescale is \citep{pz2004}
$t_{\rm rlx} \sim 10 \myr \,{} (r_{\rm hm}/0.8\pc)^{3/2}(M_{\rm TOT}/3500\msun)^{1/2}$ 
where $r_{\rm hm}$ is the half-mass radius of the YSC ($\sim 0.8-0.9\pc$ in our simulations). 
The core collapse timescale \citep{pz2002} is $t_{\rm cc}\sim 2\myr(t_{\rm rlx}/10\myr)$.

A summary of the properties of the simulated YSCs is shown in Table~\ref{tab:SCSummary}. 
These were chosen to match the properties of the most common YSCs in our Galaxy.

Each YSC was simulated for 100 Myr: at later times the YSCs are expected to be disrupted by 
the galactic tidal field (e.g. \citealt{sv2010}; \citealt{goddard2010}; \citealt{gieles2011}). 
We do not use recipes for the galactic tidal field but they will be included in future work. 
The structural evolution of our simulated YSCs is described in a companion paper (\citealt{mapelli2013}). 
From Fig.~4 of \cite{mapelli2013}, it is apparent that the half-mass radius of the YSCs at 100 Myr 
is $\sim{}3$ times the initial value. The average fraction of stars that are still bound to the 
YSC at 100 Myr is $0.85-0.9$. Thus, the simulated YSCs are expanding but most of them have not 
evaporated by the end of the simulation. This means that our results likely overestimate the 
number of dynamical exchanges and three-body encounters in the late stages of YSC life. 
We do not expect that this 
severely affects our predictions for the merger rate of BH-BH binaries, 
since the most intense dynamical activity of the YSCs occurs during (and immediately after) 
the core collapse (i.e. at $t\gtrsim{}3$ Myr), because of the dramatic increase in the core 
density (by a factor of $\ge{}10$). In fact, most of the BH-BH binaries form in the first 
$\sim{}3-40$ Myr (see the discussion in Section~\ref{sec:DBHpopulation}), and the BH-BH 
systems that are expected to merge in less than a Hubble time (and that are not disrupted 
before the end of the simulation, see Section~\ref{sec:CoalescenceTimescale}) form at $4-7$ Myr. 
In a forthcoming paper, we will add different models for the galactic tidal field, and we will 
be able to quantify their impact on the BH-BH binary population.


\section{Results}
\label{sec:results}


\subsection{DCOB population}
\label{sec:DBHpopulation}

\begin{figure}
\includegraphics[width = .9\columnwidth]{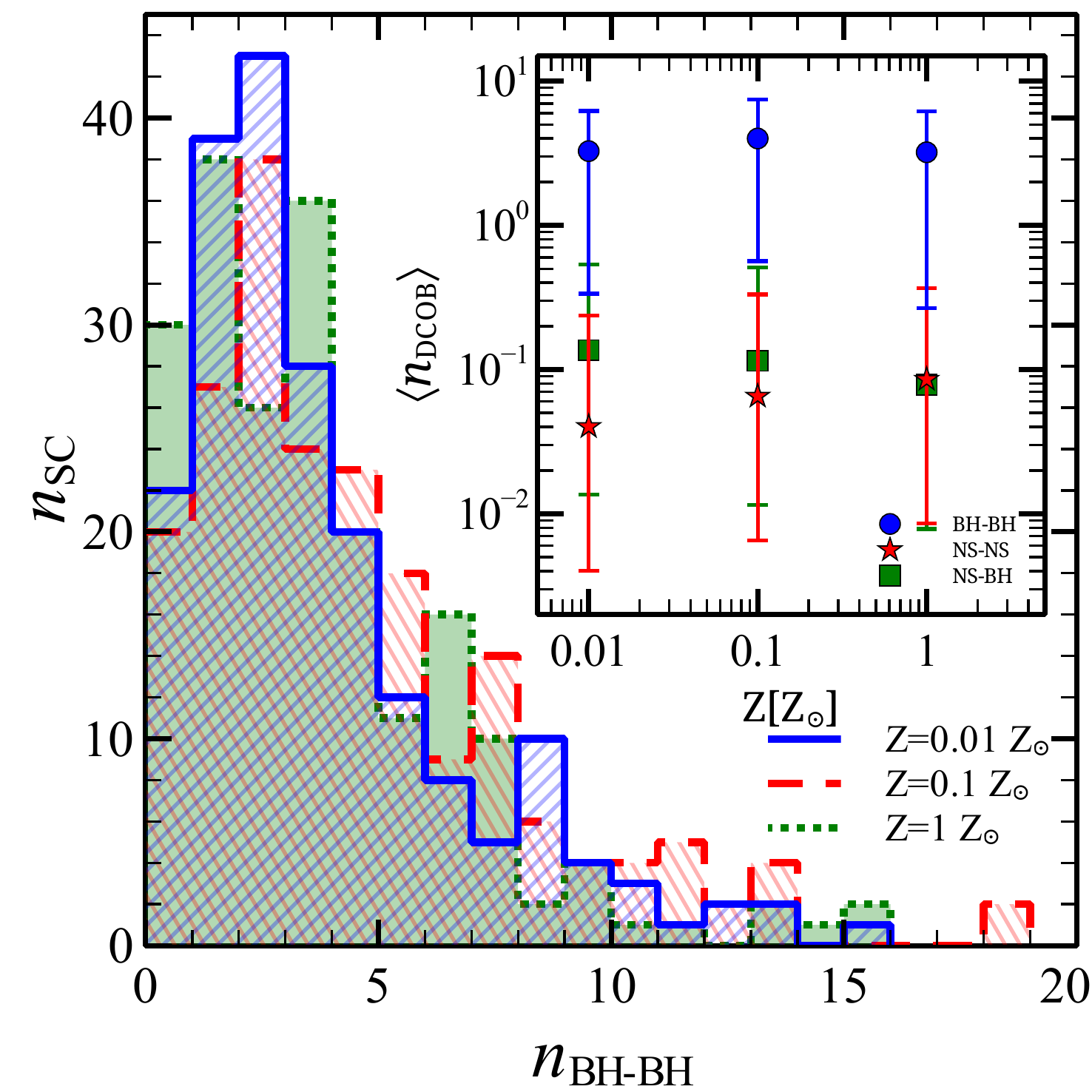}
\caption{In the main panel, distribution of the number of BH-BH binaries per YSC per metallicity. 
The blue diagonally-hatched histogram refers to $Z=0.01$ Z$_\odot$, 
the red diagonally-hatched
histogram to $Z=0.1 $Z$_\odot$ and the green filled histogram to
$Z=$ Z$_\odot$.
 In the inset, average number of BH-BH binaries (blue circles), NS-BH binaries (green squares) 
and NS-NS binaries (red stars) per YSC as a function of the YSC metallicity. The error bars are 1 $\sigma{}$ deviations.
All the quantities in this figure are integrated over the duration of the simulations (i.e. 100 Myr).}
\label{fig:meanDCOBperZ_and_distro}
\end{figure}

The number of DCOBs formed in our simulations 
is summarized in Fig.~\ref{fig:meanDCOBperZ_and_distro}. 
Here and in the following, unless otherwise specified, a binary is defined as a simulated bound pair (either existing in the initial conditions or formed during the evolution of the YSC, either hard or soft, either stable or unstable depending on the criterion adopted by {\sc starlab}, see \citealt{pz2001}). In Appendix~A, we discuss how our main results depend on this definition, by considering stable and unstable binaries separately.
 Furthermore, we classify a binary 
that forms from an exchange as a new binary with respect to the pre-exchange binary.

The inset of Fig.~\ref{fig:meanDCOBperZ_and_distro} shows that the si\-mu\-la\-ted 
number of BH-BH binaries per YSC (integrated over 100 Myr) is a factor of $\sim{}10-100$ higher than the simulated 
number of NS-NS binaries per YSC, regardless of the metallicity. 

Due to the chosen IMF, our simulated YSCs host a number of NSs that is $3-4$ 
times higher than the number of BHs. Thus, the fact that BH-BH binaries are much more 
numerous than NS-NS binaries is a striking effect of dynamics.
 BHs are heavier and tend to sink to the centre of the YSC on a timescale 
 $t_{\rm seg} \sim t_{\rm rlx}\frac{\langle m\rangle}{M_{\rm BH}}$ \citep{oleary2006}.
Thus, a 40 M$_\odot$ BH sinks towards the centre in $\sim{} 0.25$ Myr.
Once in the dense YSC centre, BHs have a higher probability to interact with 
other BHs, forming BH-BH binaries.
Furthermore,  BHs are more massive than most stars in the simulation already at 
$t\sim{}8$ Myr (when the turn-off mass is $\sim{}20$ M$_\odot$). Thus, they are 
particularly efficient in acquiring companions through dynamical exchanges 
(\citealt{hills1989}; \citealt{hills1992}).
In fact, most 
 of our BH-BH binaries come from dynamical exchanges. Only $\sim{}1.7$ per cent of BH-BH binaries 
 come from primordial binaries.
Moreover BHs have a weaker (if any) natal kick with respect to that of NSs. 
Therefore, they  are more likely to remain  in the denser regions of the YSC, rather than being ejected.

In contrast, a large fraction of NSs (up to 90 per cent at $t=100$ Myr) 
is ejected from the parent YSC as a consequence of natal kicks or dynamical recoil. The few NSs that 
remain in the YSCs are much lighter than BHs, and thus the probability that they 
acquire a second NS companion by dynamical exchanges is low. This is confirmed 
by the fact that  87 per cent  of all the NS-NS binaries come from primordial binaries.

The main panel of Fig.~\ref{fig:meanDCOBperZ_and_distro} shows the distribution 
of the number of BH-BH binaries per YSC per metallicity, integrated over the simulation time ($t_{\rm max}=100$ Myr). 
It follows a Poissonian distribution and peaks between 2 and 4 BH-BH binaries per YSC, 
in agreement with the average values shown in the inset of the same figure. 
Approximately $10-15$ per cent of YSCs do not host any 
BH-BH binary. The simulated YSC with the largest number of BH-BH binaries hosts 18 BH-BH binaries.

\begin{figure}
\includegraphics[width = \columnwidth]{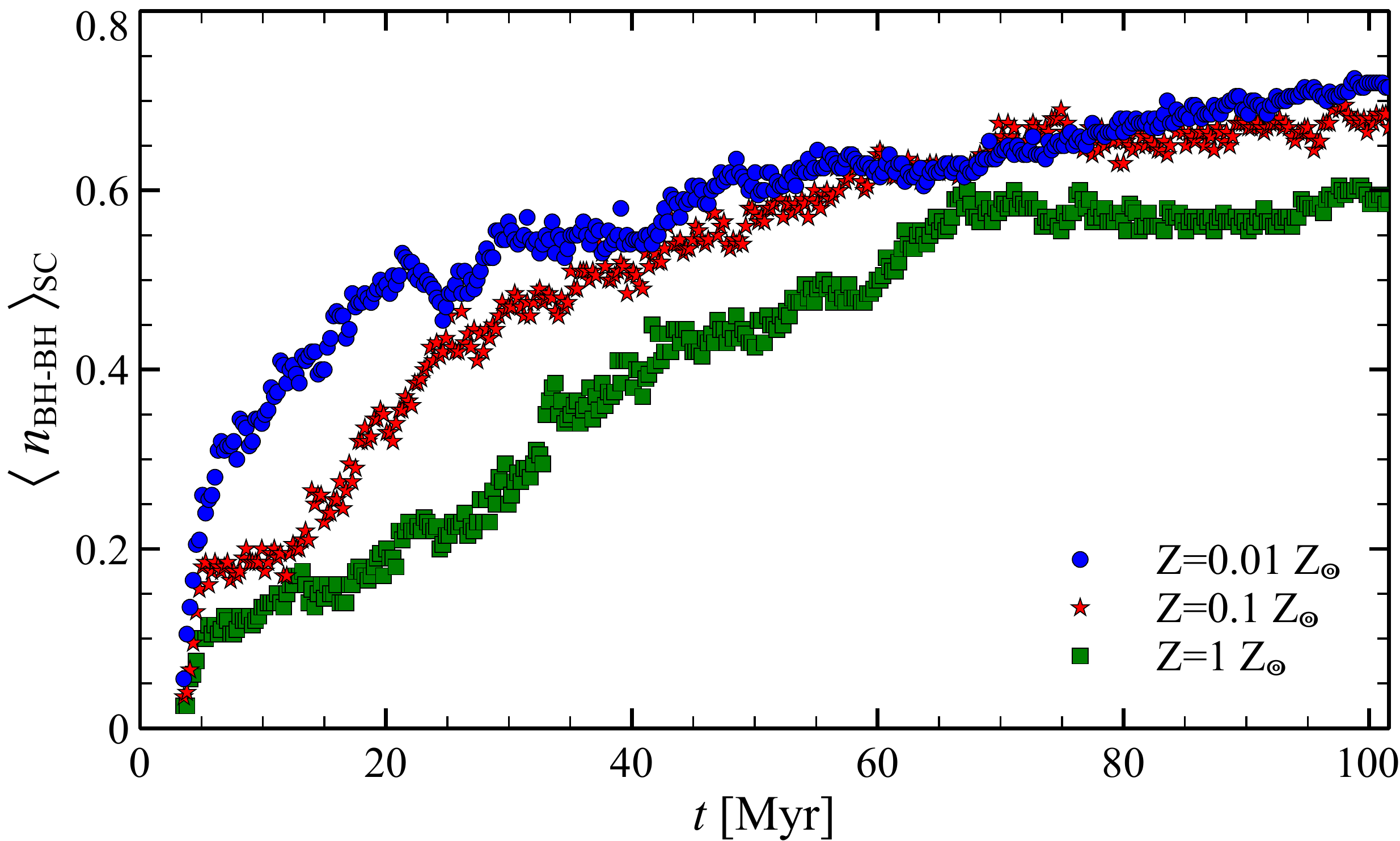}
\caption{Average number of BH-BH binaries as a function of time for the three different 
metallicities. Blue circles: $Z=0.01$ Z$_\odot$; red stars: $Z=0.1$ Z$_\odot$; 
green squares $Z=1$ Z$_\odot$. }
\label{fig:vCountInTime}
\end{figure}

We find no statistically significant differences between YSCs with different metallicity, 
when looking at the number of BH-BH binaries integrated over time (Fig.~\ref{fig:meanDCOBperZ_and_distro}). 
In contrast, we do find differences when we look at the  number of BH-BH binaries as a function of time. 
In particular, the  lower the metallicity is,  the shorter the time needed to build 
the distribution of BH-BH binaries (Fig.~\ref{fig:vCountInTime}). 

Furthermore, while in the inset of Fig. \ref{fig:meanDCOBperZ_and_distro} the average 
number of BH-BH binaries per YSC  (integrated over time) at $Z=0.1\text{ Z}_\odot$ is slightly larger than that at 
$Z=0.01\text{ Z}_\odot$,  in Fig.~\ref{fig:vCountInTime}
the  number of BH-BH binaries as a function of time at  $Z=0.01\text{ Z}_\odot$ is always 
higher than that at  $Z=0.1\text{ Z}_\odot$. This result might appear puzzling: 
the number of BH-BH binaries per YSC integrated over time is larger at 
$Z=0.1\text{ Z}_\odot$ than at $Z=0.01\text{ Z}_\odot$, while the number of 
BH-BH binaries per YSC at a given time is larger 
 at $Z=0.01\text{ Z}_\odot$ than at $Z=0.1\text{ Z}_\odot$. Actually, this is a 
 consequence of the fact 
that BHs are more massive at low metallicity, 
and thus are more efficient in acquiring companions through dynamical exchanges
and in producing stable binaries with longer lifetimes. This implies that the  
BH-BH binaries which form at $Z=0.01\text{ Z}_\odot$ are less numerous than those 
which form at $Z=0.1,\,{}1\text{ Z}_\odot$ but they live for a much longer time 
(before being ionized or exchanged) than the latter (see Fig.~\ref{fig:lifetimes} 
and the comments in next section). Thus, if we look at a YSC at a given time, we 
find more BH-BH binaries at $Z=0.01\text{ Z}_\odot$ than at $Z=0.1,\,{}1\text{ Z}_\odot$.

Finally, we notice that the first BH-BH binaries form at $t\sim{}3$ Myr, i.e. the time 
of core collapse, regardless of the metallicity. This is a consequence of the fact 
that binary hardening becomes important during the core collapse and drives the 
re-expansion of the core \citep{mapelli2013}.

\subsection{Lifetimes and exchanges}
\label{sec:LifetimesAndExchanges}

\begin{figure}
 \includegraphics[width = \columnwidth]{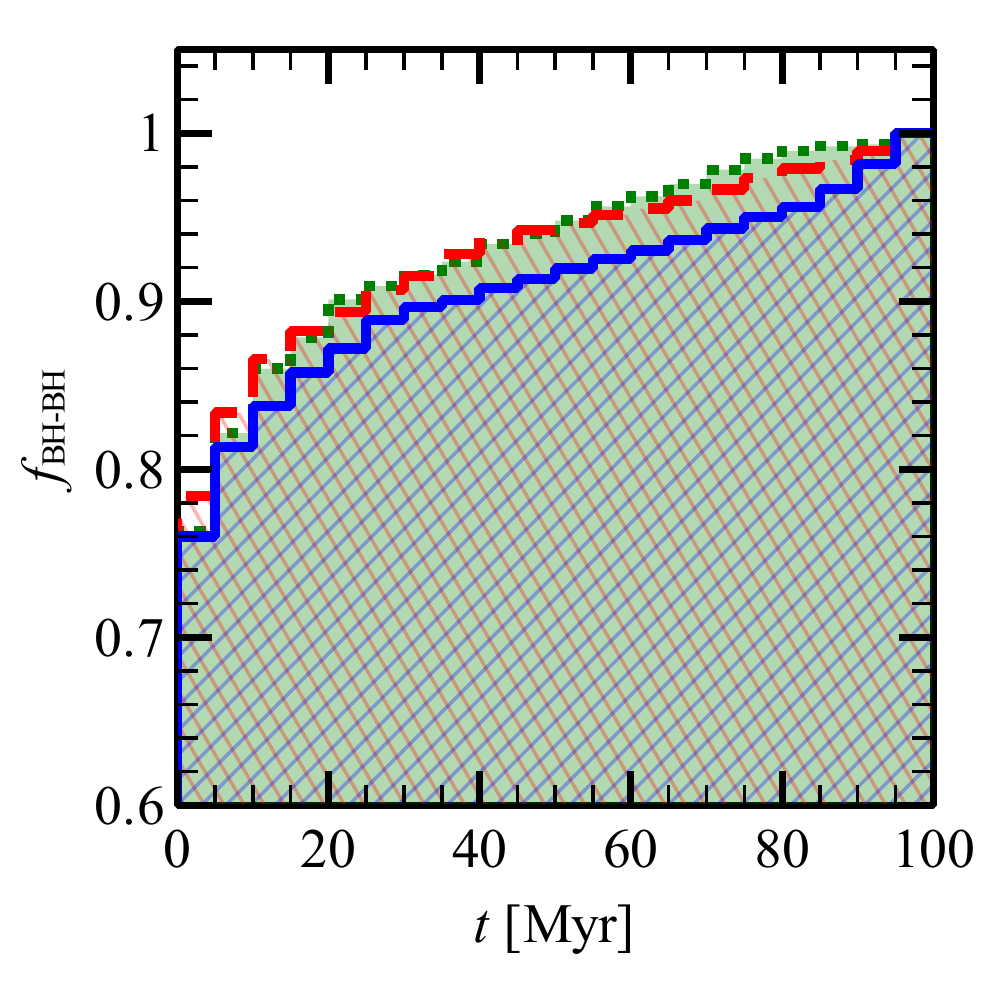}
\caption{ Cumulative distribution of BH-BH binary lifetimes (normalized to the 
total number of BH-BH binaries per each metallicity).  
Blue diagonally-hatched histogram: $Z=0.01$ Z$_\odot$; red diagonally-hatched 
histogram: $Z=0.1$ Z$_\odot$; green filled histogram: $Z=$ Z$_\odot$. }
\label{fig:lifetimes}
\end{figure}

In section \ref{sec:DBHpopulation}, we showed that metal-poor YSCs 
build up their BH-BH binary population earlier than metal-rich ones. 
Furthermore, the BH-BH binaries that form in metal-poor YSCs ($Z=0.01$ Z$_\odot$) are 
more stable, i.e. have longer lifetimes (before they break up or undergo another 
exchange). This is a consequence of the higher BH masses allowed 
in the failed SN scenario. 
In Fig.~\ref{fig:lifetimes}, we show the cumulative 
distribution of BH-BH binary lifetimes.
At $Z=0.1, 1\text{ Z}_\odot$ 90 per cent of BH-BH binaries survive for less 
than 20 Myr, while at $Z=0.01\text{ Z}_\odot$ 90 per cent of BH-BH binaries survive up 
to 40 Myr. Furthermore, about 5 per cent of BH-BH binaries survive for more than 80 Myr 
in the YSCs with $Z=0.01\text{ Z}_\odot$, while only 1--2 per cent of BH-BH binaries 
survive for more than 80 Myr in the YSCs with $Z\ge{}0.1\,{}{\rm Z}_\odot$.


We have also run a Kolmogorov-Smirnov (KS) test on the distributions presented in 
Fig.~\ref{fig:lifetimes}. We find a probability $P_{\rm KS} = 4.05\times 10^{-8}$ that 
BH-BH binary lifetimes at $Z = 0.01\text{ Z}_\odot$ and at $Z = 0.1\text{ Z}_\odot$ are 
drawn from the same distribution. Similarly, $P_{\rm KS} = 5.46\times 10^{-2}$ when comparing 
BH-BH binary lifetimes at $Z = 0.01\text{ Z}_\odot$ and $Z = \text{Z}_\odot$, and 
$P_{\rm KS} = 3.14\times 10^{-6}$  when comparing BH-BH binary lifetimes at $Z = 0.1\text{ Z}_\odot$ and $Z = \text{Z}_\odot$. 
This result confirms that the three distributions are statistically different.


We notice that the average number 
of exchanges is quite the same across different metallicities in 
Table~\ref{tab:avgExchangesTable}. 
Thus, the difference in lifetimes must be interpreted as
a higher probability of binary break up (i.e. ionization) in case of high 
metallicity. 
Also, from Table~\ref{tab:avgExchangesTable} we notice  that the few survived 
NS-NS binaries are very stable, as they undergo a low number of exchanges.

\begin{figure*}
 \includegraphics[width = \textwidth]{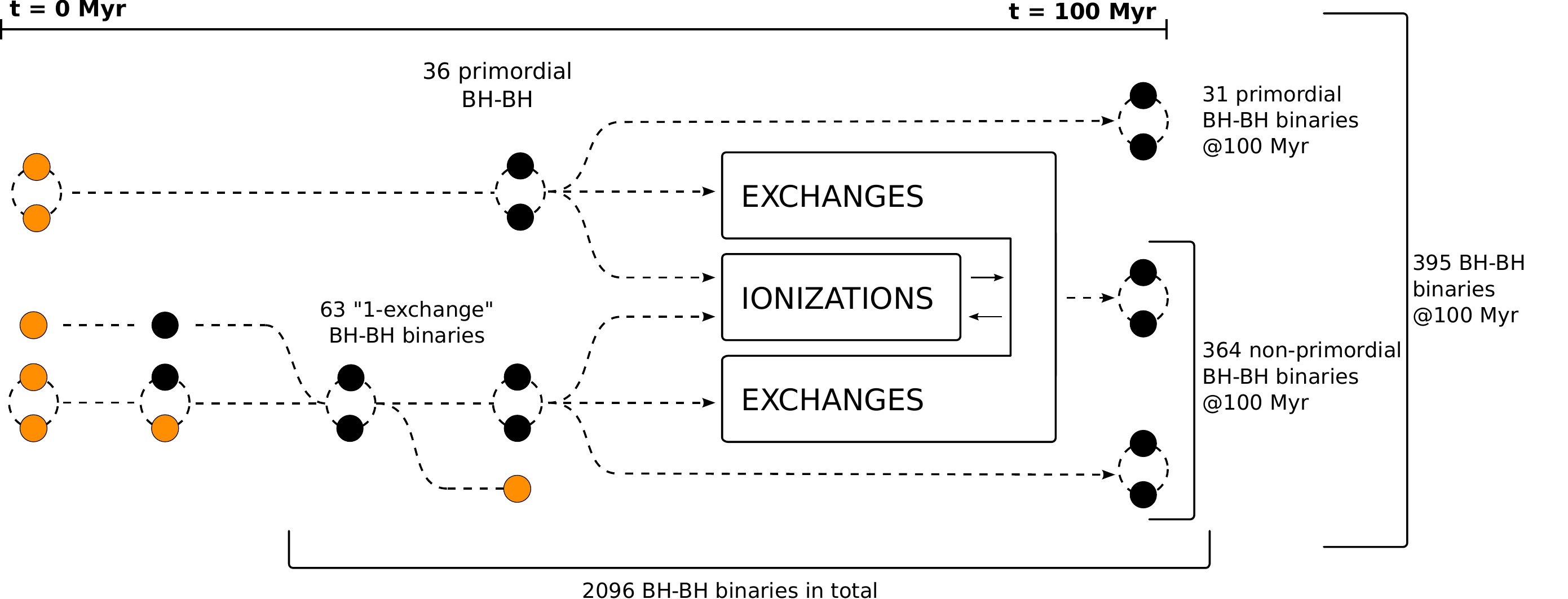}
\caption{Schematic representation of the main formation and evolution pathways of 
BH-BH binaries in our simulations. Yellow circles: stars; black circles: black holes. 
In the top row, from left to right: primordial binaries can evolve into BH-BH binaries 
by stellar evolution. Then, primordial BH-BH binaries can be ionized or undergo an 
exchange and become exchanged BH-BH binaries.
In the bottom row,  from left to right: we call '1-exchange' binaries those BH-BH 
binaries that form after the exchange of a BH into a BH-star binary. In the following, 
'1-exchange' binaries can either be ionized or undergo more exchanges. For the sake of 
simplicity, we call ionizations also the exchanges that transform a BH-BH binary into a 
BH-star binary. The members of a ionized BH-BH binary can enter a BH-BH binary again via 
three-body exchange.}
\label{fig:exchStats}
\end{figure*}

Fig. \ref{fig:exchStats} summarizes the possible pathways that lead to the 
formation of a BH-BH binary and their relative importance in our simulations. 
BH-BH binaries can derive from either a primordial binary or an exchange. 
The upper branch of the scheme shows that 36 simulated BH-BH binaries are primordial binaries, 
while 63 simulated BH-BH binaries form through a dynamical exchange in which a single 
BH replaces a star in a BH-star binary (in Fig. \ref{fig:exchStats}, these systems are 
called '1-exchange' BH-BH binaries).

In the subsequent evolution, BH-BH binaries born from primordial binaries can either 
be ionized by a three-body encounter, or undergo an exchange. If the primordial binary 
undergoes an exchange and if the intruder is a BH, the BH-BH binary becomes an 
exchanged BH-BH binary.
%
Considering the entire set of simulations for 100 Myr, the total number of 
BH-BH binaries formed is 2096.

At the end of the simulations (i.e. after 100 Myr) the BH-BH binaries that still 
survive are 31 primordial binaries and 364 exchanged binaries, for a total of 395 
BH-BH binaries (0.66 BH-BH binaries per YSC, on average).

Thus, in summary,  1.7 per cent of all BH-BH binaries in our simulations are primordial 
binaries, while the remaining 97.3 per cent are exchanged binaries.



 \begin{table}
 \caption{Average number of exchanges per metallicity per DCOB type. Values 
 outside (within parenthesis) refer to all DCOBs (only DCOBs that are considered
 `stable' according to the criterion defined in  {\sc StarLab}, see \citealt{pz2001} and our Appendix~A).}
 \begin{tabular}{lccc}
\hline
Type & 0.01 Z$_\odot$ & 0.1 Z$_\odot$ & Z$_\odot$ \\
\hline
BH-BH & 9.92 (0.41) & 9.91 (0.48) & 10.14 (0.58)\\
NS-NS & 0.00 (0.00) & 0.50 (0.15) &  0.26 (0.09)\\
NS-BH  & 6.33 (0.49) & 3.72 (0.48) &  3.48 (0.43)
\vspace{0.01cm}\\
\hline
\end{tabular}
\label{tab:avgExchangesTable}
\end{table}


\subsection{Orbital properties}\label{sec:OrbitalProperties}
\begin{figure*}
\includegraphics[width = 16cm]{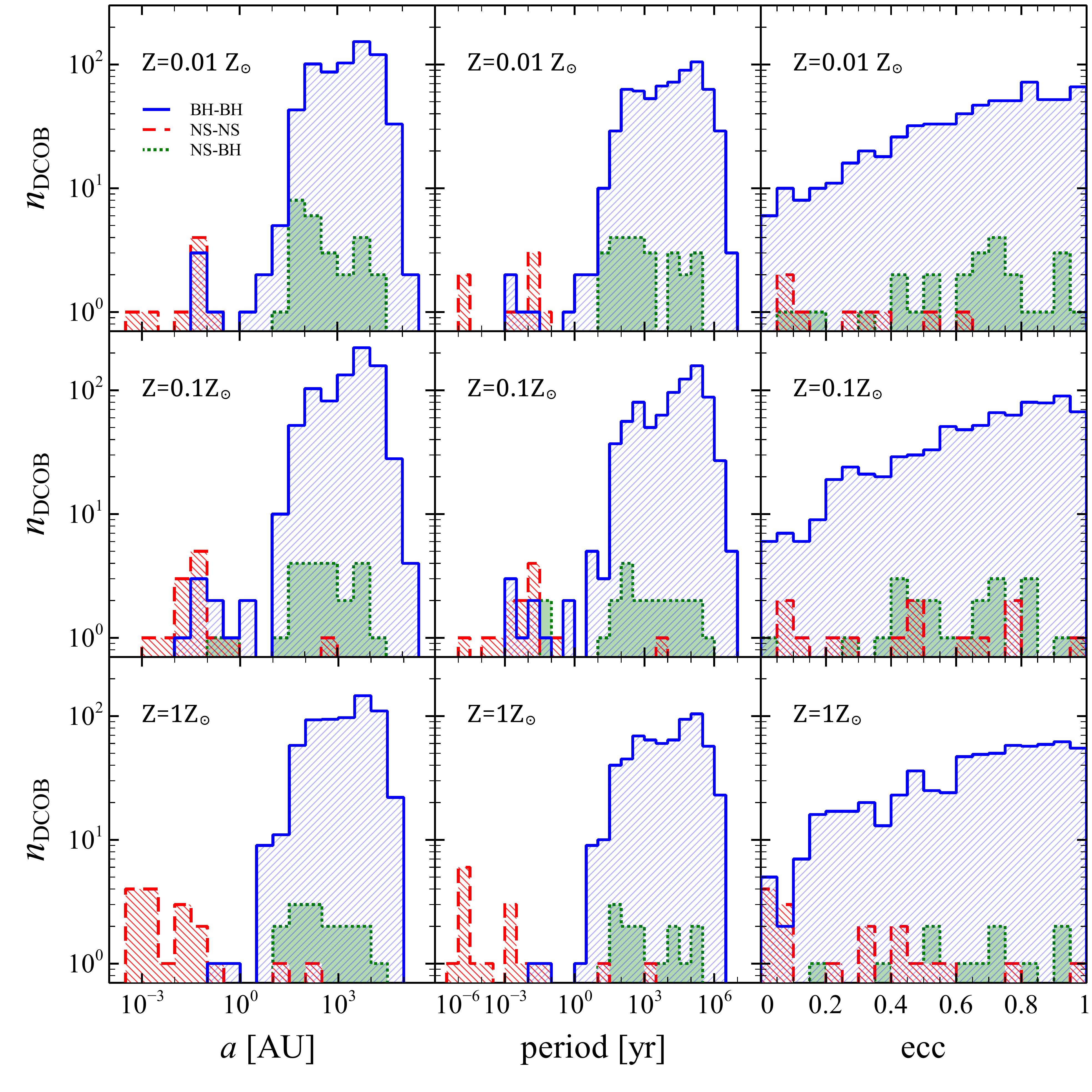}
\caption{Histograms of the orbital properties of DCOBs measured, for each binary, 
when the semi-major axis $a$ is minimum. 
Columns from left to right refer to semi-major axis $a$, 
period and eccentricity of the binary. Rows  from top to bottom refer 
to three different metallicities: 
$Z=0.01,\,{}0.1$ and  $1\,{}{\rm Z}_\odot$.  The blue, red and green histograms 
refer to BH-BH, NS-NS and NS-BH binaries, respectively.}
\label{fig:vOrbPropHist}
\end{figure*}

In Fig.~\ref{fig:vOrbPropHist}, the distributions of the orbital properties of 
the BH-BH, NS-NS and NS-BH binaries are shown. These are measured at the time in which the 
semi-major axis $a$ is minimum for each binary.
%
The metallicity does not significantly affect the distribution of semi-major axes and eccentricities of BH-BH binaries.
The eccentricity distribution of BH-BH binaries follows the initial 
equilibrium distribution $f(e)\propto 2e$, but with 
an excess of low-eccentricity systems coming from the circularization by tidal 
forces (which influenced some systems before both components collapsed) and by GW emission.

BH-BH binaries span a wide range in both semi-major axes and orbital periods 
($10^{-2}-10^6\text{ AU}$ and $10^{-3}-10^7$ yr, respectively).

We notice a strong break in the distribution of semi-major axes of BH-BH binaries at $\sim 1$ AU. 
This is consistent with the fact that the most massive primordial binaries with 
separation $a\lesssim 1$ AU merged before the formation of BHs, 
emptying the region of BH-BH binaries with that semi-major axis.
Only dynamical effects can populate this region, but they do it slowly, 
because the hardening time (i.e. the timescale for hardening a binary by 
three-body encounters) scales as $a^{-2}$ (see e.g. \citealt{quinlan1996}).

We notice that the softest binaries in Fig.~\ref{fig:vOrbPropHist} have semi-major axis as large as $\sim{}5$ pc, close to the initial tidal radius of the YSC. These extremely loose bound pairs are highly unstable (see the discussion in the appendix) and very short-lived: it is reasonable to expect that they would completely disappear, if a galactic tidal field would be included in our simulations.

NS-NS binaries are much less numerous than BH-BH binaries (as we 
showed in Figure~\ref{fig:meanDCOBperZ_and_distro}), but the distribution of 
their orbital parameters indicates that NS-NS binaries have generally smaller semi-major 
axes than BH-BH binaries. This may be due to a selection effect: as NS-NS binary progenitors are 
often ionized either by natal kicks or by exchanges involving more massive 
stellar objects (e.g. BHs), only the hardest NS-NS binaries survive in our simulations.

Finally, NS-BH binaries are about 10 times less numerous than BH-BH binaries, 
but follow approximately the same distribution of orbital parameters.


\subsection{Mass distribution}
\label{sec:MassDistribution}

\begin{figure}
 \includegraphics[width = \columnwidth]{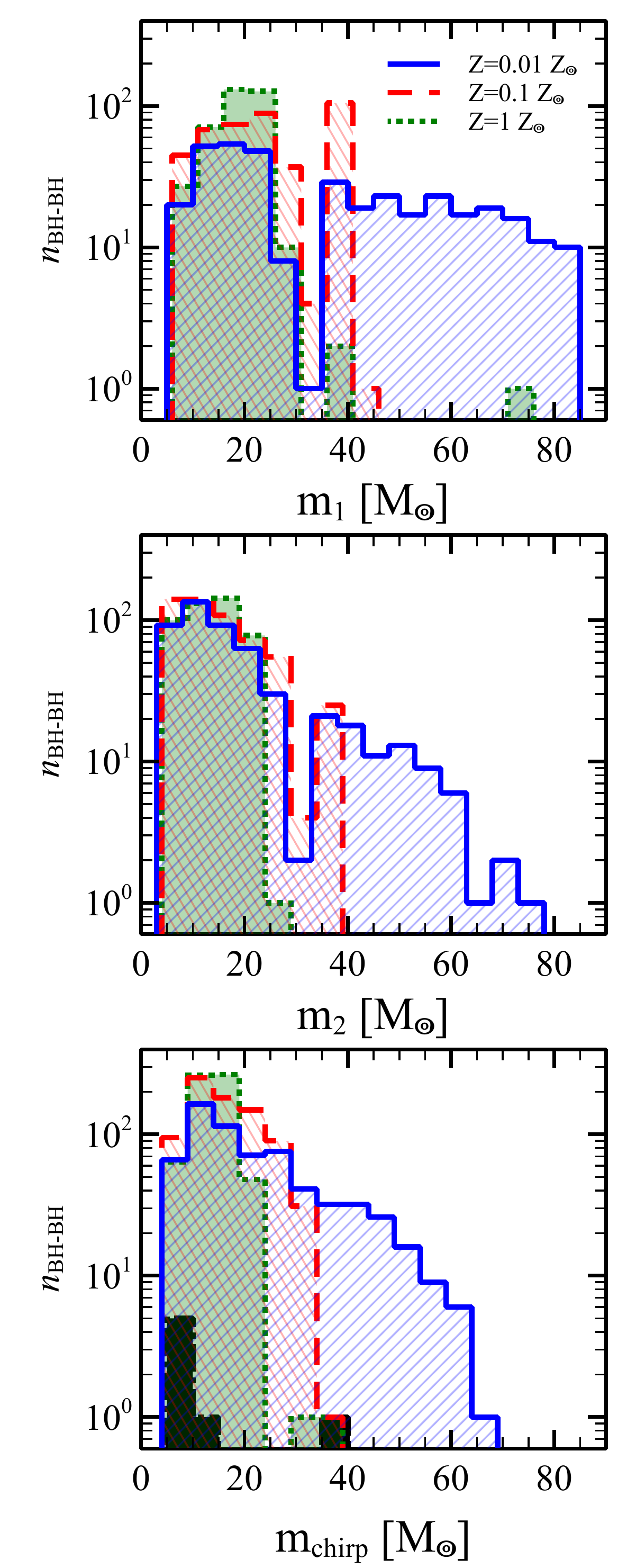}
\caption{From top to bottom: distribution of the primary component, of the 
secondary component and of the chirp mass of BH-BH binaries, respectively. In each 
panel, the blue, red and green histograms correspond to $Z=0.01,\,{}0.1,\,{}1\,{}{\rm Z}_\odot{}$, respectively. 
In the bottom panel, the black histograms show the distribution of chirp 
masses of the 7 BH-BH binaries that are expected to merge within a Hubble time 
(see Section~\ref{sec:CoalescenceTimescale}).}
\label{fig:vMassHist}
\end{figure}

The mass of the BHs affects both the frequency and the amplitude of 
the GW signal (e.g. \citealt{maggiore2008}). Thus, it is important to look at 
the distribution of the masses of the simulated BH-BH binaries.

Fig.~\ref{fig:vMassHist} shows the distribution of $m_1$, $m_2$ and of the 
chirp mass $m_{\rm chirp}$. The chirp mass is defined as 
$m_{\rm chirp}=(m_1\,{}m_2)^{3/5}/(m_1+m_2)^{1/5}$.  The chirp mass is named so because
it is this combination of $m_1$ and $m_2$ that determines how fast the binary sweeps, or chirps, through
a frequency band. In fact, it can be shown that the 
amplitude and the frequency of GWs scale as $m_{\rm chirp}^{5/3}$ and 
$m_{\rm chirp}^{-5/8}$, respectively (\citealt{maggiore2008}).

The mass of the primary (secondary) can be as high as $85 \msun$ ($78 \msun$) 
in case of $Z=0.01$ Z$_\odot$. Such  large values  correspond to BHs that formed 
from direct collapse (see Section \ref{sec:MethodsAndSimulations}
and paper~I).

We also found a $73 \msun$ BH at $Z=\text{ Z}_\odot$, i.e. a much higher 
mass than expected from stellar evolution of isolated stars with solar metallicity. 
This BH is the result 
of a dynamically induced merger between a smaller BH (14.9 M$_\odot$)
 and a star (59.3 M$_\odot$).

Chirp masses are very high, too. The black histogram in Fig.~\ref{fig:vMassHist} 
shows the chirp mass distribution of our best BH-BH merger candidates (i.e. of those 
systems that are expected to merge within a Hubble time, see next section for 
details): we notice that one of these systems has a significantly high chirp 
mass ($m_{\rm chirp} \simeq 40\,$M$_\odot$).

The GW searches for BH-BH binaries performed by LIGO and VIRGO \citep{abadie2012a, aasi2013} cover the mass range found
by the present simulation. The signal corresponding to our higher chirp masses can be detected by
the Intermediate Mass Binary Black Holes search \citep{abadie2012b}.

In the adopted model, the chirp mass strongly depends on the metallicity of 
the progenitor stars. Since the amplitude and the frequency of GWs scale as 
$m_{\rm chirp}^{5/3}$ and $m_{\rm chirp}^{-5/8}$, respectively, it will be 
possible to link the observed GW signal to the chirp mass of the source. Observing large chirp masses
would be clear evidence for the scenario of BH birth and evolution in the low metallicity
environments.


\subsection{Coalescence timescale}
\label{sec:CoalescenceTimescale}

\begin{figure*}
\includegraphics[width = \textwidth]{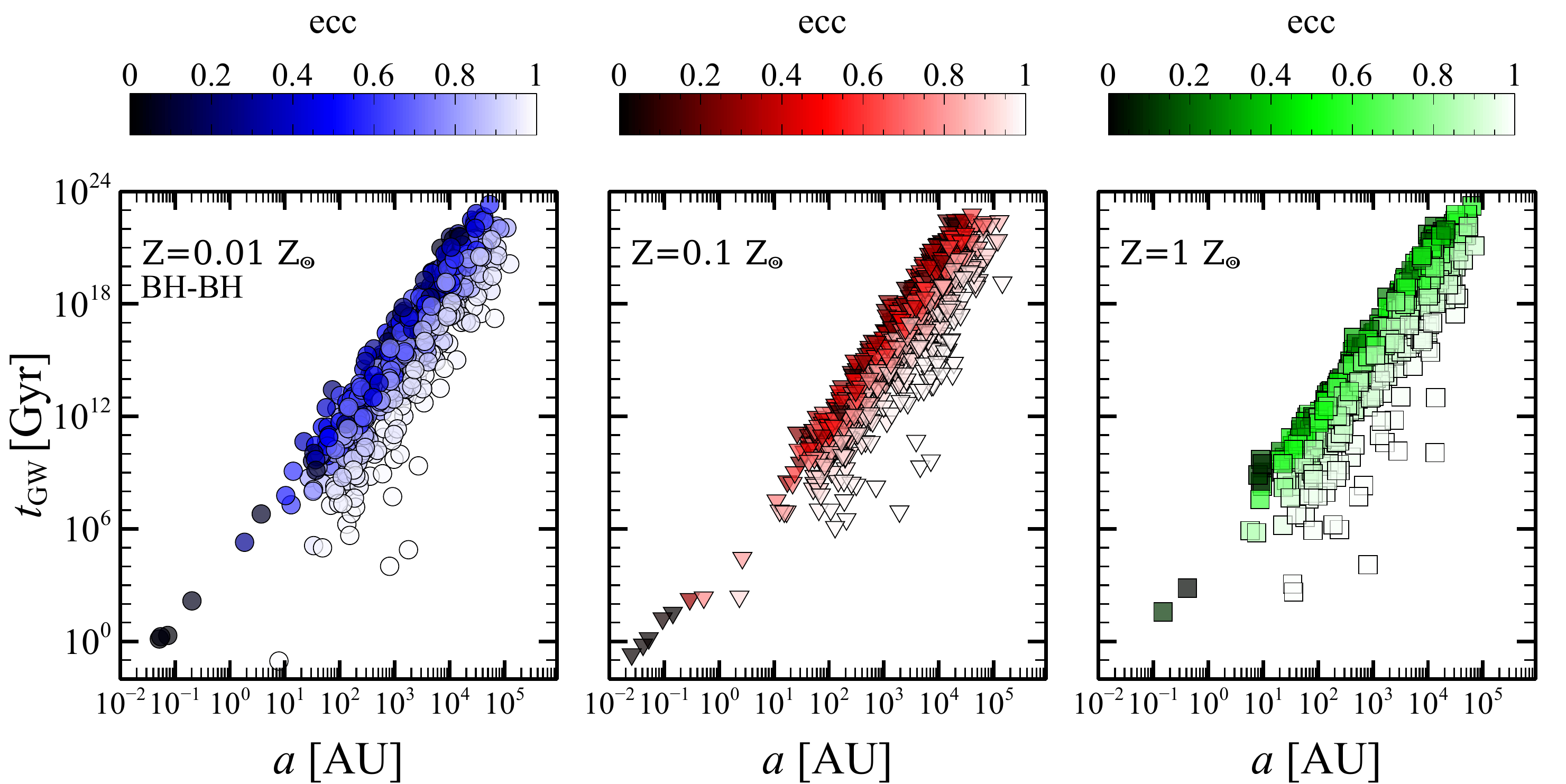}\\
\vspace{-2cm}
\includegraphics[width = \textwidth]{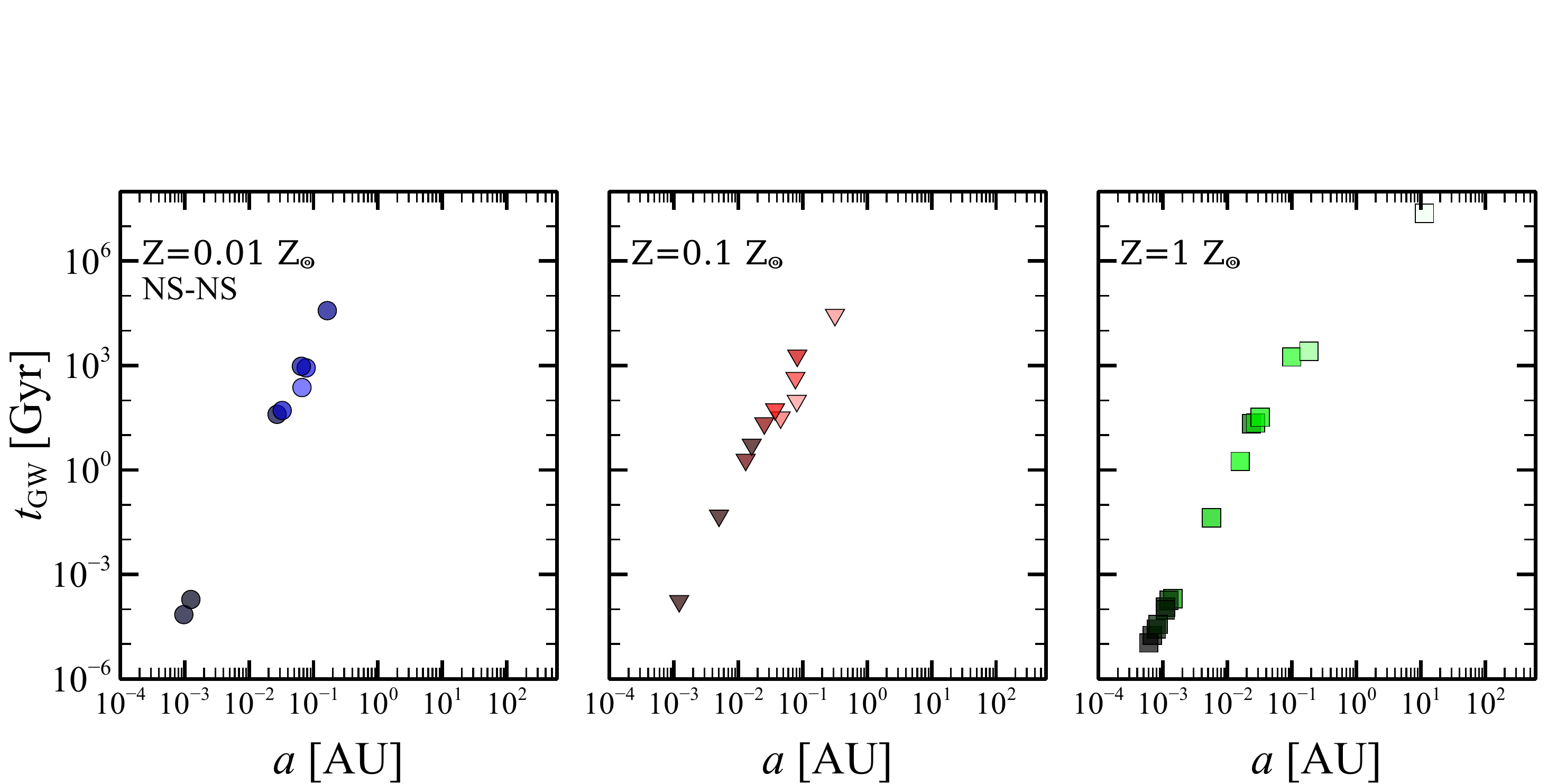}\\
\vspace{-2cm}
\includegraphics[width = \textwidth]{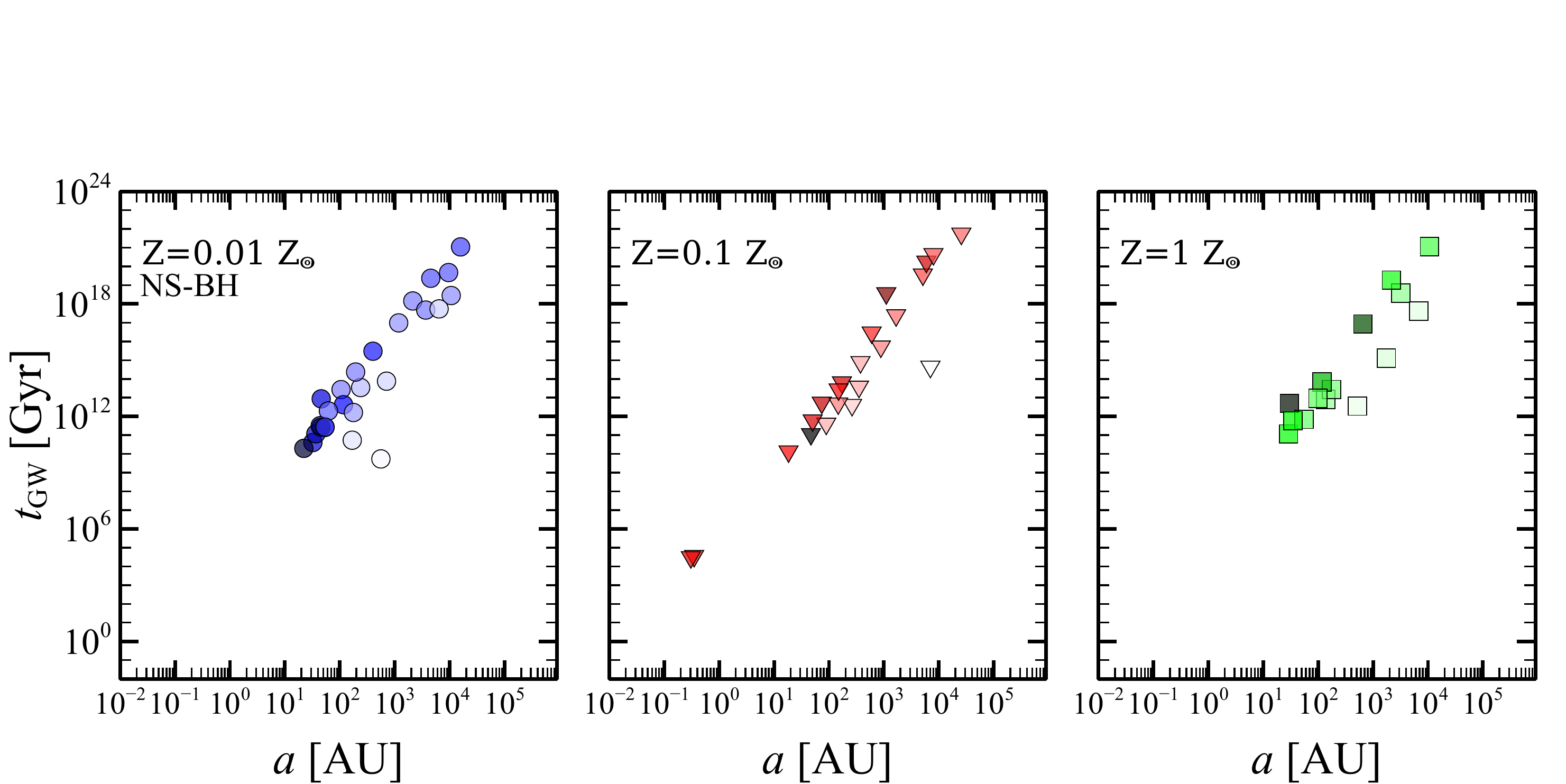}
\caption{Coalescence timescale as a function of the semi-major axis for BH-BH binaries, 
 NS-NS binaries and NS-BH from top to bottom.
From left to right: metallicity $Z=0.01$ Z$_\odot$ (blue circles), 0.1 Z$_\odot$ 
(red  triangles) and 1 Z$_\odot$ (green squares). The colour-coded map refers to 
eccentricity. }
\label{fig:vPetersTimesMap} 
\end{figure*}

\begin{table*}
 \begin{tabular}{rrrrrr}
\hline
   $t_{\rm GW}$ (Gyr) &  $a$ (AU) & $P$ (yr) & ecc & $Z$ (Z$_\odot$) & Merger \\
    \hline
    0.09 & 7.77 & 2.27788 & 0.997 & 0.01 & N \\
    0.20 & 0.03 & 0.00107 & 0.019 & 0.1  & N \\
    0.67 & 0.04 & 0.00196 & 0.019 & 0.1  & N \\
    1.34 & 0.05 & 0.00267 & 0.019 & 0.01 & N \\
    1.49 & 0.05 & 0.00276 & 0.014 & 0.1  & N \\
    1.76 & 0.05 & 0.00296 & 0.028 & 0.01 & N \\
    2.06 & 0.07 & 0.00387 & 0.016 & 0.01 & N \\
    \hline
  \end{tabular}
  \caption{List of the BH-BH binaries with coalescence timescale $<13$ Gyr, 
  in ascending order of coalescence timescale. Column 1: coalescence timescale 
  in Gyr; column 2: semi-major axis in AU; column 3: period in years; column 4: 
  eccentricity; column 5:  metallicity; column 6: whether or not (Y/N) the binary 
  merges during the simulation.}
  \label{tab:DBHpetersTable}
 \end{table*}
 
 \begin{table*}
 \begin{tabular}{rrrrrr}
\hline
    $t_{\rm GW}$ (Myr) &  $a$ ($10^{-3}$ AU) & $P$ ($10^{-5}$ yr) & ecc & $Z$ 
    (Z$_\odot$) & Merger \\
    \hline
	0.01        & 0.60    & 0.9  & 0.005 & 1 & Y\\
	0.02        & 0.69    & 1.1 & 0.01 & 1 & Y\\
	0.03        & 0.79    & 1.4 & 0.05  & 1 & Y\\
	0.04        & 0.84    & 1.5 & 0.05  & 1 & Y\\
	0.07        & 0.97    & 1.9   & 0.08  & 0.01 & Y\\
	0.1        & 1.09    & 2.2 & 0.06  & 1 & Y \\
	0.1        & 1.1    & 2.4 & 0.03  & 1 & Y\\
	0.2        & 1.2    & 2.7    & 0.09  & 0.1 & Y\\
	0.2        & 1.2    & 2.7   & 0.08  & 1 & Y\\
	0.2        & 1.3    & 2.8  & 0.06  & 0.01 & Y\\
	0.2        & 1.4    & 3.3    & 0.31   & 1 & Y\\
	40         & 5.7    & 26.9  & 0.42   & 1 & N \\
	50         & 5.0    & 22.3  & 0.09  & 0.1 & N \\
	1760       & 20     & 12.5  & 0.51   & 1 & N \\
	1960       & 10     & 93.4  & 0.21   & 0.1 & N \\
	5330       & 20     & 13.0  & 0.11   & 0.1 & N \\
    \hline         
  \end{tabular}    
  \caption{List of NS-NS binaries with coalescence time $<13$ Gyr,  in ascending order of 
  coalescence timescale. Column 1: coalescence timescale in Myr; column 2: 
  semi-major axis in units of $10^{-3}$ AU; column 3: period in units of 
  $10^{-5}$ yr; column 4: eccentricity; column 5:  metallicity; column 6: 
  whether or not (Y/N) the binary merges during the simulation. The minimum, 
  mean and maximum difference between the real merger and the coalescence 
  times are 0.02, 0.24 and 0.12 Myr, respectively. }
  \label{tab:DNSpeters&Table}
 \end{table*}
 
The timescale for coalescence \citep{peters1964} is defined as 
\begin{equation}
t_{\rm GW} = \frac{5}{256}\frac{c^5\,{}a^4\,{}(1-e^2)^{7/2}}{G^3\,{}m_1\,{}m_2\,{}(m_1+m_2)},
\end{equation}
where $c$ is the speed of light and $G$ the gravitational constant. $t_{\rm GW}$ is 
the timescale for a binary to merge by GW emission.  It scales as $a^4$, and it 
is shorter for high eccentricity. 
GW emission affects the coalescence timescale by shrinking the semi-major axis 
and circularizing the binary orbit. Fig.~\ref{fig:vPetersTimesMap} 
shows $t_{\rm GW}$ as a function of semi-major axis, eccentricity and metallicity of the simulated systems. 

Most of the systems with $t_{\rm GW}\le{}t_{\rm H}$ (where $t_{\rm H}=13$ Gyr is 
the Hubble time) have eccentricity close to zero, as a consequence of 
circularization by GW emission. However, we found an outlier 
(with eccentricity $e=0.997$, see Table \ref{tab:DBHpetersTable}) 
produced by dynamical exchange. This is 
  interesting not only because its coalescence timescale is short, 
  due to the high value   of the eccentricity, but also because it 
  suggests 
  that the use of templates which include eccentric effects in the LIGO and
VIRGO searches could be important \citep{brown+2010}.  
  Unfortunately, this binary is destroyed by a new dynamical exchange 
  before it merges.   On the other hand, we expect to find other 
  systems like this with a larger simulation 
  sample, and we cannot exclude that some of them can evolve (without 
  being destroyed by   further exchanges) till they merge. Such systems would be 
  very important for GW detection  \citep{brown+2010, samsing+2013}.


All BH-BH binaries with $t_{\rm GW}\le{}t_{\rm H}$ are at low metallicity ($Z=0.01$ Z$_\odot$ and 
$Z=0.1$  Z$_\odot$), while we find none at solar metallicity.
 The bottom panels in Fig.~\ref{fig:vPetersTimesMap} show the coalescence 
 timescale for NS-NS binaries. The total number of NS-NS binaries is much smaller than that 
 of BH-BH binaries but 
 they are much harder. As a consequence, their coalescence timescales are 
 generally shorter. The minimum coalescence timescale for BH-BH binaries in our simulations 
 is $t_{\rm GW}\sim 0.1$ Gyr, while that  for NS-NS binaries is $t_{\rm GW}\sim 10^{-5}$ Gyr. 
 We also found that 11 NS-NS binaries actually merged before 100 Myr.  

In the bottom panel of Fig.~\ref{fig:vPetersTimesMap}, we also show the coalescence 
timescale for NS-BH binaries. No NS-BH mergers are expected in less than a Hubble time, 
because NS-BH binaries are much less numerous than BH-BH binaries and they are not favoured by dynamical encounters.

Tables \ref{tab:DBHpetersTable} and  \ref{tab:DNSpeters&Table} list the shortest 
coalescence timescales for BH-BH binaries and NS-NS binaries, respectively. 


It has been debated (e.g. \citealt{clausen2012}) whether $ t_{\rm GW}$ is a 
reliable indicator of the merger timescale in star clusters. 
In fact, dynamical interactions in star clusters may affect the evolution of a 
DCOB and delay or anticipate the merger with respect to the expected $ t_{\rm GW}$. 
 In our simulations there is good agreement between the coalescence timescales and the actual mergers, 
thus, we can conclude that in most cases dynamics does not affect the actual merger 
timescale of the simulated NS-NS binaries.




\section{Discussion}
\label{sec:Discussion}


\subsection{Estimate of the merger rate}
%
%
%
Since most stars form in YSCs, the mass density of YSCs in the Universe is expected to scale as the star formation rate (SFR) density (\citealt{mapelli2010}). 
Thus, from the results discussed in Section~\ref{sec:CoalescenceTimescale} and using a Drake-like equation, the merger rate of BH-BH binaries can be estimated as 
\begin{eqnarray}\label{eq:mrgrDBHB}
R_{\rm BH-BH}=N_{\rm mrgr,\,{}BH-BH}\quad{}\rho_{\rm SF}\quad{}t_{\rm life}\quad{}f_{\rm SF}\nonumber\\=3.5\times{}10^{-3} \,{}{\rm Mpc}^{-3}\,{}{\rm Myr}^{-1}\nonumber\\\left[\frac{N_{\rm mrgr,\,{}BH-BH}}{3\times{}10^{-15}\,{}{\rm M}_\odot{}^{-1}\,{}{\rm yr}^{-1}}\right]\,{}\left(\frac{\rho_{\rm SF}}{1.5\times{}10^{-2}\,{}{\rm M}_\odot{}\,{}{\rm yr}^{-1}\,{}{\rm Mpc}^{-3}}\right)\nonumber\\\left(\frac{t_{\rm life}}{10^8{\rm yr}}\right)\,{}\left(\frac{f_{\rm SF}}{0.8}\right),
\end{eqnarray}
where $\rho_{\rm SF}$ is the cosmological density of SFR at redshift zero 
($\rho_{\rm SF}=1.5\times{}10^{-2}\,{}{\rm M}_\odot{}\,{}{\rm yr}^{-1}\,{}{\rm Mpc}^{-3}$ 
from \citealt{hopkins2006}), $t_{\rm life}$ is the average lifetime of a YSC, 
$f_{\rm SF}$ is the fraction of star formation (SF) that occurs in YSCs 
(we take $f_{\rm SF}=0.8$ from \citealt{ladalada2003}), and $N_{\rm mrgr,\,{}BH-BH}$ 
is the number of BH-BH binary mergers per solar mass per year, as estimated from our 
simulations (see Table~\ref{tab:DBHpetersTable}). In equation~\ref{eq:mrgrDBHB}, we assume that $R_{\rm BH-BH}$ does not change significantly with time. This approximation is reasonable for the distance range of Advanced LIGO and VIRGO (see the short discussion at the end of this section).

Equation~\ref{eq:mrgrDBHB} has been derived following the same approach as explained in Mapelli et al. (2010a, see also \citealt{mapelli2012a}). The main differences between equation~\ref{eq:mrgrDBHB} of this paper and equations 2 and 3 of \cite{mapelli2010} are the following: (i) in equation~\ref{eq:mrgrDBHB} we just estimate the merger rate, while \cite{mapelli2010} estimate the detection rate for different interferometers; (ii) in equation~\ref{eq:mrgrDBHB} we derive $N_{\rm mrgr,\,{}BH-BH}$ directly from our simulations, while in \cite{mapelli2010} we used the results of a toy model for intermediate-mass BHs.

In particular, we estimate $N_{\rm mrgr,\,{}BH-BH}$ as
\begin{eqnarray}\label{eq:mrgrDBHB2}
N_{\rm mrgr,\,{}BH-BH}=3\times{}10^{-15}{\rm M}_\odot{}^{-1}\,{}{\rm yr}^{-1}\left(\frac{N_{\rm exp,\,{}BH-BH}}{3}\right)\,{}\nonumber\\\left(\frac{200}{N_{\rm YSC}}\right)\,{}\left(\frac{3500\,{}{\rm M}_\odot{}}{\langle{}M_{\rm TOT}\rangle{}}\right)\,{}\left(\frac{1.5\,{}{\rm Gyr}}{t_{\rm GW,\,{}max}}\right),
\end{eqnarray}
where $N_{\rm YSC}$ is the number of simulated YSCs, $\langle{}M_{\rm TOT}\rangle{}$ is the average mass of a single YSC\footnote{Since we simulated only YSCs with $M_{\rm TOT}\sim{}3500$ M$_\odot{}$, equation~\ref{eq:mrgrDBHB2} suffers from the approximation that we do not consider a mass spectrum for the simulated YSCs. On the other hand, YSCs with $M_{\rm TOT}\sim{}3500$ M$_\odot{}$ are among the most diffuse YSCs in the local Universe (\citealt{ladalada2003}). In a forthcoming paper, we will consider a mass spectrum for the YSCs.}, and $N_{\rm exp,\,{}BH-BH}$ is the number of BH-BH binaries that are expected to merge within a time $t_{\rm GW,\,{}max}$. For example, at $Z=0.1$ Z$_\odot$, we find that 3 BH-BH binaries are expected to merge within $t_{\rm GW,\,{}max}=1.5$ Gyr (see Table~\ref{tab:DBHpetersTable}).  At $Z=0.01$ Z$_\odot$, we find  4 BH-BH binaries are expected to merge within $t_{\rm GW,\,{}max}=2.1$ Gyr, while at $Z=1$ Z$_\odot$ we do not find any BH-BH binaries that merge within $t_{\rm GW,\,{}max}=t_{\rm H}$. Thus, we find that $0\le{}N_{\rm mrgr,\,{}BH-BH}\le{}3\times{}10^{-15}{\rm M}_\odot{}^{-1}\,{}{\rm yr}^{-1}$ depending on the metallicity. The resulting values of the merger rate are $R_{\rm BH-BH}=0$, $3.3$, and $3.5\times{}10^{-3} \,{}{\rm Mpc}^{-3}\,{}{\rm Myr}^{-1}$, if we assume that all YSCs in the local Universe have metallicity $1,\,{}0.1$ and 0.01 Z$_\odot{}$, respectively.


Thus, the merger rate of BH-BH binaries is $R_{\rm BH-BH}\sim{}3.5\times{}10^{-3} \,{}{\rm Mpc}^{-3}\,{}{\rm Myr}^{-1}$ if we assume that all YSCs in the local Universe formed at low metallicity ($Z\le{}0.1\,{}{\rm Z}_\odot{}$), and is $R_{\rm BH-BH}\sim{}0$ if we assume that all YSCs in the local Universe formed at high metallicity ($Z={\rm Z}_\odot{}$), since in our simulations we did not find any BH-BH binary at $Z={\rm Z}_\odot{}$ with coalescence timescale shorter than the Hubble time. 
Even if the statistics is low, this result is important, as  
we can conclude that BH-BH binaries are enhanced at low metallicity, where more massive BHs can form. 

As a first-order approximation, we can assume that the merger rate of BH-BH binaries in the local Universe is included in this range of values, i.e. $0\le{}R_{\rm BH-BH}\le{}3.5\times{}10^{-3} \,{}{\rm Mpc}^{-3}\,{}{\rm Myr}^{-1}$. For a more realistic assumption about the metallicity of YSCs in the local Universe, see the discussion at the end of this Section.

Similarly, the merger rate of NS-NS binaries can be estimated as
\begin{eqnarray}\label{eq:mrgrDNSB}
R_{\rm NS-NS}=N_{\rm mrgr,\,{}NS-NS}\quad{}\rho_{\rm SF}\quad{}t_{\rm life}\quad{}f_{\rm SF}\nonumber\\=0.15\,{}{\rm Mpc}^{-3}\,{}{\rm Myr}^{-1}\nonumber\\\left[\frac{N_{\rm mrgr,\,{}NS-NS}}{1.3\times{}10^{-13}\,{}{\rm M}_\odot{}^{-1}\,{}{\rm yr}^{-1}}\right]\,{}\left(\frac{\rho_{\rm SF}}{1.5\times{}10^{-2}\,{}{\rm M}_\odot{}\,{}{\rm yr}^{-1}\,{}{\rm Mpc}^{-3}}\right)\nonumber\\\left(\frac{t_{\rm life}}{10^8{\rm yr}}\right)\,{}\left(\frac{f_{\rm SF}}{0.8}\right),
\end{eqnarray}
where $N_{\rm mrgr,\,{}NS-NS}$ is the number of NS-NS binary mergers per solar mass per year and can be derived as
\begin{eqnarray}\label{eq:mrgrDNSB2}
N_{\rm mrgr,\,{}NS-NS}=1.3\times{}10^{-13}{\rm M}_\odot{}^{-1}\,{}{\rm yr}^{-1}\left(\frac{N_{\rm exp,\,{}NS-NS}}{9}\right)\,{}\nonumber\\\left(\frac{200}{N_{\rm YSC}}\right)\,{}\left(\frac{3500\,{}{\rm M}_\odot{}}{\langle{}M_{\rm TOT}\rangle{}}\right)\,{}\left(\frac{100\,{}{\rm Myr}}{t_{\rm life}}\right),
\end{eqnarray}
where $N_{\rm exp,\,{}NS-NS}$ is the number of NS-NS binaries that actually merged during our simulations and  $t_{\rm life}=100$ Myr is the assumed YSC life (and the duration of the simulation). In the case of NS-NS binaries we use the number of merged binaries (rather than the number of expected mergers, as in the case of BH-BH binaries), because we have sufficient statistics to do so. At $Z=1,\,{}0.1$ and 0.01 Z$_\odot$ $N_{\rm exp,\,{}NS-NS}=9,\,{}2,\,{}2$, respectively.

Thus, the merger rate of NS-NS binaries is $R_{\rm NS-NS}\sim{}0.15 \,{}{\rm Mpc}^{-3}\,{}{\rm Myr}^{-1}$  if we assume that all YSCs in the local Universe formed at high metallicity ($Z={\rm Z}_\odot{}$, see Table~\ref{tab:DNSpeters&Table}), and is $R_{\rm NS-NS}\sim{}0.03\,{}{\rm Mpc}^{-3}\,{}{\rm Myr}^{-1}$ if we assume that all YSCs in the local Universe formed low metallicity ($Z=0.01,\,{}0.1\,{}{\rm Z}_\odot{}$). This is another important 
results of our simulations, as it implies that NS-NS mergers are suppressed 
at low metallicity.

As a first-order approximation, we can assume that the merger rate of NS-NS binaries in the local Universe is included in this range of values, i.e. $0.03\,{}{\rm Mpc}^{-3}\,{}{\rm Myr}^{-1}\le{}R_{\rm NS-NS}\le{}0.15\,{}{\rm Mpc}^{-3}\,{}{\rm Myr}^{-1}$. In equation~\ref{eq:mrgrDNSB}, we assume that $R_{\rm NS-NS}$ does not change significantly with time. This approximation is reasonable for the distance range of Advanced LIGO and VIRGO (see the short discussion at the end of this Section).

Finally, the merger rate of NS-BH binaries  is 
$R_{\rm NS-BH}<10^{-4}\,{}{\rm Mpc}^{-3}\,{}{\rm Myr}^{-1}$ for all considered 
metallicities, as we found no simulated systems with coalescence timescale 
shorter than the Hubble time. In our simulations, NS-BH systems are much less 
common than BH-BH binaries, since the latter are favoured by dynamical exchanges with 
respect to the former.

Our estimates of the merger rate show that there is a possible trend with metallicity:
the mergers of NS-NS binaries are favoured at high metallicity ($\sim{}{\rm Z}_\odot{}$), while the mergers of BH-BH binaries are more frequent
at low metallicity ($\sim{}0.01-0.1$ Z$_\odot{}$). We recall that $Z=0.01$  Z$_\odot{}$ is the typical metallicity of GCs in the Milky Way (e.g. \citealt{harris96}), $Z=0.1$  Z$_\odot{}$ is the metallicity of many irregular galaxies and dwarf galaxies in the local Universe  (e.g. \citealt{mapelli2010b}), while a metallicity close to solar is normally found in the bulges of giant spiral galaxies and elliptical galaxies (e.g. \citealt{pilyugin2004}).  Furthermore, a metallicity gradient (with $Z$ decreasing at larger distance from the centre) has been found in most local late-type galaxies (\citealt{pilyugin2004}). Thus, the metallicity of the local Universe is quite patchy, with a preference for higher metallicity at the centre of the most massive galaxies and for lower metallicity in the outskirts of massive galaxies as well as in dwarf and irregular galaxies. 

Furthermore, the Sloan Digital Sky Survey shows that the SF in the last Gyr has a bimodal distribution: about half of it occurs at solar metallicity, while the remaining half takes place at $Z\sim{}0.1$ Z$_\odot{}$ (\citealt{panter2008}). Therefore, we expect that about half of the YSCs that formed in the last Gyr have $Z\sim{}$ Z$_\odot$, while the remaining half have $Z\sim{}0.1$ Z$_\odot$. In contrast, a negligible fraction of YSCs formed at $Z=0.01$ Z$_\odot{}$ in the last Gyr. 

If we assume (as suggested by \citealt{panter2008}) that half of the YSCs that formed in the last Gyr have $Z\sim{}$ Z$_\odot$, while the remaining half have $Z\sim{}0.1$ Z$_\odot$, the rate of mergers we expect today from our simulated YSCs  (using equations \ref{eq:mrgrDNSB} and \ref{eq:mrgrDBHB}) is $R_{\rm NS-NS}\sim{}0.10 \,{}{\rm Mpc}^{-3}\,{}{\rm Myr}^{-1}$  and $R_{\rm BH-BH}\sim{}1.7\times{}10^{-3} \,{}{\rm Mpc}^{-3}\,{}{\rm Myr}^{-1}$, for NS-NS and BH-BH binaries, respectively.

The aforementioned values of $R_{\rm NS-NS}$ and $R_{\rm BH-BH}$ have been derived from the typical properties of YSCs in the local Universe and assuming a metallicity mixture valid for the last Gyr (i.e. up to redshift $z\sim{}0.1$). Are they valid over the entire distance range of Advanced LIGO and VIRGO? According to \cite{abadie2010}, the distance range of Advanced LIGO and VIRGO will be $\sim{}200$ Mpc ($z\sim{}0.05$) and 1 Gpc ($z\sim{}0.2$) for NS-NS and BH-BH mergers, respectively. Thus, we can conclude that our estimated merger rates are fairly uniform (within the uncertainties) across the range of Advanced LIGO and VIRGO, especially in the case of NS-NS mergers.

We recall that 
the DCOBs that form in YSCs will be ejected to the field as a consequence of 
evaporation, natal kicks  and three-body encounters, and because of the 
disruption of the parent YSCs by the tidal field of the host galaxy. Thus, the 
merger rate we estimate in this Section represents the expected merger rate for 
the field. 
This is very important, as previous studies estimated the merger rate either for 
for long-lived GCs (e.g. \citealt{oleary2006}; \citealt{downing2010}; 
\citealt{downing2011}) or for the field (e.g. \citealt{b2010a}; \citealt{dominik2012}; 
\citealt{dominik2013}). In previous work, the effect of dynamics has been included 
only in the estimate of the merger rate within GCs, while field binaries have been 
assumed to form and evolve in isolation (through population synthesis codes). 
On the other hand, it is well known that most stars form in YSCs and evolve 
dynamically via three-body encounters, before being ejected into the field. 
Our results show that the estimate of the merger rate in the field should account for dynamical evolution.

\subsection{Comparison with previous work}


\begin{figure}
\includegraphics[width = \columnwidth]{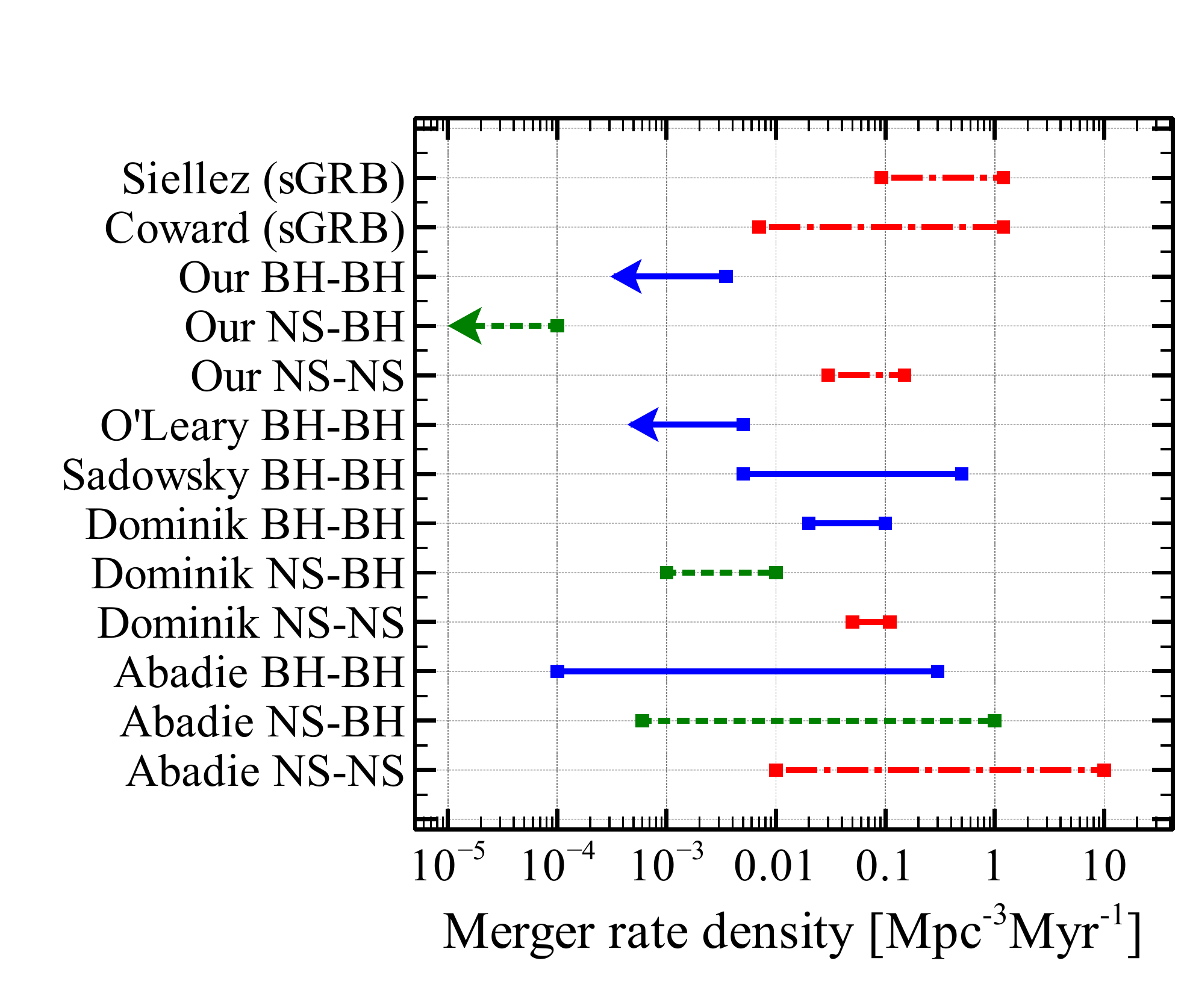}
\caption{Comparison of our predictions for the merger rates of NS-NS, NS-BH and 
BH-BH binaries with some of the most representative estimates available in the literature. From top to bottom: \protect\cite{siellez2014}; \protect\cite{coward2012}; our paper; \protect\cite{oleary2006}; \protect\cite{sadowski2008}; \protect\cite{dominik2013}; \protect\cite{abadie2010}.
 The predicted merger rates for \protect\cite{dominik2013} span from their ``Standard'' to their ``Optimistic CE'' model (see Fig. 1 in \protect\cite{dominik2013}).}
\label{fig:mgrComparison} 
\end{figure}

Fig.~\ref{fig:mgrComparison} compares our predictions of the merger rates with some of the most representative estimates available in the literature. From this Figure, it is apparent that our prediction of $R_{\rm NS-NS}$ is fairly consistent with the estimate derived from short gamma-ray bursts (\citealt{coward2012, siellez2014}). 

Furthermore, our results for $R_{\rm NS-NS}$ and $R_{\rm BH-BH}$ are consistent with the estimates provided in \citet{abadie2010}. In contrast, our 
results for $R_{\rm NS-BH}$ are significantly lower than predicted by \citet{abadie2010}. 
We recall that the value of $R_{\rm NS-NS}$ reported by \citet{abadie2010} is 
derived from the observed rate of NS-NS binaries in the Milky Way (\citealt{kalogera2004}), 
while the values of $R_{\rm NS-BH}$ and $R_{\rm BH-BH}$ are obtained from population 
synthesis codes (\citealt{oshaughnessy2008} and \citealt{kalogera2007}, respectively) 
and are only indirectly constrained by the SN rate. 

The main differences between the 
approach presented in \citet{abadie2010} and ours are the following. (i) 
The estimates presented in  \citet{abadie2010} are based on population synthesis 
simulations of isolated binaries and do not account for the fact that most stars 
form in YSCs; (ii) the mass spectrum of BHs is significantly different; (iii) 
\citet{abadie2010} assume that most galaxies in the local Universe are Milky Way 
analogues, while in this paper we adopt the cosmic SFR by \citet{hopkins2006}. 
The fact that we account for the dynamical evolution of YSCs and include more massive 
BHs than \citet{abadie2010} affects the results significantly, as the formation of 
BH-BH binaries is enhanced with respect to that of NS-BH systems. In general, our simulated 
DCOBs cannot evolve in isolation but frequently undergo three-body encounters that 
perturb their orbits, while the results of \citet{abadie2010} are obtained assuming 
that all binaries evolve in isolation.

Recent studies by \citet{b2010a}, \citet{dominik2012} and \citet{dominik2013} adopt 
a BH mass spectrum much more similar to ours and investigate the dependence of the 
merger rate on metallicity, even if they do not include three-body encounters. 
As a consequence, the distribution of BH-BH binary chirp masses in the three aforementioned 
papers is very similar to our distribution. The main difference is the absence of 
massive BHs that come from a merger in the papers by \citet{b2010a}, 
\citet{dominik2012} and \citet{dominik2013}, because they do not allow merged 
binaries to acquire a new companion dynamically. 


In their standard model, \citet{dominik2013} find an estimate of $R_{\rm NS-NS}$  that is fairly consistent with ours, while their prediction for $R_{\rm NS-BH}$ and $R_{\rm BH-BH}$ are about a factor of ten higher. In addiction, \citet{b2010a}, \citet{dominik2012} and \citet{dominik2013} present  an alternative model in which  common envelope (CE) phases on the Hertzsprung gap 
are allowed (i.e. the binary is not assumed to merge when one of the two members 
reaches the Hertzsprung gap). The merger rates obtained with this assumption are 
a factor of $\ge{}100$ higher than our results. This discrepancy is consistent 
with our expectations, as our simulations adopt the same recipes for the CE phase 
as in the standard model of \citet{dominik2013}\footnote{As discussed in paper~I, we adopt $\alpha{}_{\rm CE}\,{}\lambda{}=0.5$ to model the CE phase (see \citealt{davis2012} for a definition), and we assume that all binaries that enter a CE phase when at least one of the two members is in the Hertzsprung gap merge.}.

\citet{sadowski2008} study the merger rate of DCOBs in GCs and in the field by 
means of Monte Carlo simulations and population synthesis models, respectively.
They find that NS-NS binaries and NS-BH binaries should dominate the DCOB population in 
the field, whereas BH-BH binaries are the main merger candidates in GCs. We confirm their 
result, in the sense that the formation of BH-BH binaries is enhanced by dynamics in star 
clusters. Our results agree with those of \citet{sadowski2008} also for the 
importance of dynamical exchanges: \citet{sadowski2008} find that 6 per cent 
(94 per cent) of BH-BH binary merger candidates come from primordial binaries (dynamical 
exchanges), while we find that 1.7 per cent of our BH-BH binaries come from primordial 
binaries. 

On the other hand, \citet{sadowski2008} neglect the fact that many of the merger 
candidates in the field have been ejected from YSCs (by dynamical ejection, natal 
kick or YSC disruption). Accounting for field DCOBs that were ejected from YSCs 
increases the relative importance of BH-BH binaries in the field, especially at low metallicity. 
Furthermore, \citet{sadowski2008} find a merger rate  $R_{\rm BH-BH}\sim{}0.005-0.5$ 
Mpc$^{-3}$ Myr $^{-1}$ in dense star clusters, substantially higher than our result 
($R_{\rm BH-BH}\le{}0.0035$ Mpc$^{-3}$ Myr $^{-1}$), because they assume that the BHs 
remain in dynamical equilibrium with the rest of the cluster.  This suppresses the 
dynamical ejection of BHs.

Other recent papers (\citealt{oleary2006}; \citealt{downing2010}; \citealt{downing2011}; 
\citealt{clausen2012}) focus on DCOB merger in dense stellar systems and GCs. 
In particular, \citet{oleary2006} perform Monte Carlo simulations of GCs in which 
they assume that the BH population is concentrated in the core and dynamically 
decoupled from the rest of the cluster, because of the Spitzer instability 
(\citealt{spitzer1969}). \citet{oleary2006} find that most BH-BH binaries are ejected 
from the parent star cluster and that the resulting merger rate is 
$R_{\rm BH-BH}\le{}0.005$ Mpc$^{-3}$ Myr $^{-1}$, much lower than in 
\citet{sadowski2008}, because of the assumed Spitzer instability. 
The merger rate estimated by \citet{oleary2006} is very similar to our result. 

 \citet{downing2010} and \citet{downing2011} perform Monte Carlo simulations of GCs. 
 They (i) include a treatment of metallicity that is close to ours (even if their 
 maximum BH mass is generally lower than ours, as they use the same distribution 
 as in \citealt{b2006}), (ii)  assume neither Spitzer instability nor rigid 
 equilibrium between the BHs and the rest of the cluster {\it a priori}.  
 \citet{downing2010} find that the BHs strongly mass segregate and evolve 
 similarly to what assumed by \citet{oleary2006}. \citet{downing2010} find an 
 even lower merger rate than  the one derived by \citet{oleary2006} and by our 
 paper, but they admit that this may be due to their approximate treatment of 
 three-body encounters. 
On the other hand, the distribution of orbital periods  in the simulations by 
\citet{downing2010} is similar to ours (see Fig.~\ref{fig:vOrbPropHist}). Furthermore, 
both this paper and \citet{downing2010} find that most BH-BH binaries form dynamically, 
through exchanges. Finally, \citet{downing2010} find that BH-BH binaries form earlier and 
are more stable at low metallicity, because BHs are more massive, in agreement 
with our results (see Figures \ref{fig:vCountInTime} and \ref{fig:lifetimes}). 

In conclusion, our results confirm that most BH-BH binaries in star clusters come from 
dynamical exchanges, in  agreement with the findings of Monte Carlo simulations 
of dense star clusters (\citealt{oleary2006}; \citealt{downing2010}; 
\citealt{downing2011}). On the other hand, our simulated star clusters are a 
factor of $10-1000$ less massive and a factor of $\ge{}5$ smaller than those 
studied in previous work (e.g. \citealt{downing2010}). Thus, they are expected 
to be much more numerous in the local Universe than those considered by previous 
work (since the mass function of YSCs scales as $M_{\rm TOT}^{-2}$, 
\citealt{ladalada2003}). Furthermore, the dynamical evolution timescale of our 
simulated YSCs is much shorter, as $t_{\rm rlx}\sim{}10\,{}{\rm Myr}\,{}(r_{\rm hm} /0.8\,{} {\rm pc})^{3/2}\,{}(M_{\rm TOT} /3500\,{}{\rm M}_\odot{})^{1/2}$.
  Thus, most DCOBs that form in our simulated YSCs will be ejected to the field 
  (by YSC evaporation, three-body encounters or tidal fields), over a 
  timescale much shorter than found in previous work. Therefore, our YSCs can 
  be considered the  building blocks of the galaxy disc, and the merger rate we 
  have estimated represents the expected merger rate of the field population.


\section{Conclusions}
\label{sec:Conclusions}
We studied the impact of metallicity and dynamics on the formation and evolution of
DCOBs. To this purpose, we have run 600 N-body realizations of YSCs chosen to match the 
properties of the most common YSCs in our Galaxy. We simulated YSCs, because most 
stars form in YSCs. Thus, we cannot 
study the formation and evolution of DCOBs without accounting for the fact that most of them originate in YSCs.

For our simulations, we used an upgraded version of the public code {\sc starlab}, which 
includes recipes for metallicity-dependent 
stellar evolution and winds, and which allows stars with final
 mass larger than $40\, M_\odot$ to directly collapse to a BH. Direct collapse leads 
 to the formation of massive stellar BHs ($\geq 25\,M_\odot$) at low metallicity.

 We found that, while the number of NSs is about four times larger than the number 
 of BHs, the number of BH-BH binaries is about ten times higher than the number of NS-NS binaries.  
 The reason is that dynamical interactions enhance the formation of BH-BH binaries with 
 respect to NS-NS binaries. Heavier BHs sink to the centre of the YSC, where they are more 
 likely to interact with other BHs: BHs can acquire companions
through three-body exchanges. Since the probability of a dynamical exchange is 
higher when the single star is more massive than one of the members of the 
binary and since BHs are among the most massive objects in a YSC, exchanges 
favour the formation of BH-BH binaries.

BH-BH binaries form earlier at low metallicity, because BHs are more massive in metal-poor YSCs. 
Furthermore, BH-BH binaries formed at low metallicity are more stable: they live longer than 
BH-BH binaries in metal-rich YSCs.

The simulated BH-BH binaries have very large chirp masses ($5-70$ M$_\odot$), because of the direct collapse at 
low metallicity and because mergers between stars and BHs are allowed. 

BH-BH binaries span a wide range in periods ($10^{-3}-10^7$ yr). In contrast, most NS-NS binaries
have periods $<1$ yr. As a consequence, the coalescence timescale is generally 
longer for BH-BH binaries than for NS-NS binaries. The minimum coalescence timescale for BH-BH binaries and 
NS-NS binaries is $t_{\rm GW}\sim 0.1$ Gyr and $t_{\rm GW}\sim 10^{-5}$ Gyr, respectively.
Only 7 BH-BH binaries are expected to merge within a Hubble time. Moreover, no BH-BH binaries merge 
during our simulations, while 11 NS-NS binaries do.

From our simulations, we can estimate the merger rate of DCOBs in the local Universe. 
We find a merger rate $R_{\rm BH-BH}\le{}3.5\times{}10^{-3} \,{}{\rm Mpc}^{-3}\,{}{\rm Myr}^{-1}$, $R_{\rm NS-BH}<10^{-4}\,{}{\rm Mpc}^{-3}\,{}{\rm Myr}^{-1}$ and 
$R_{\rm NS-NS}\sim{}0.03-0.15 \,{}{\rm Mpc}^{-3}\,{}{\rm Myr}^{-1}$ 
for BH-BH, NS-BH and NS-NS binaries, respectively. The merger rate of NS-NS binaries is fairly consistent 
with the estimates based on both the observed Galactic NS-NS binaries (\citealt{kalogera2004}) and the observed rate of short gamma-ray bursts (\citealt{coward2012, siellez2014}). 
The merger rate of BH-BH binaries is consistent with recent Monte Carlo simulations of dense 
star clusters (e.g. \citealt{oleary2006}; \citealt{downing2010}). The merger rate 
of NS-BH binaries is quite low with respect to previous estimates based on 
population synthesis codes (e.g. \citealt{oshaughnessy2008}). This can be 
explained with the fact that the formation of NS-BH binaries is less favoured 
by dynamical exchanges than the formation of BH-BH binaries.

Our merger rates are still affected by a number of assumptions that will be 
improved in forthcoming studies. First, in our study we assume that the lifetime 
of the simulated YSCs is  100 Myr, but we do not take into account the presence 
of a realistic galactic tidal field. Second, we explore only a limited portion 
of the parameter space. In forthcoming studies, we will consider YSCs with 
different concentration, half-mass radius, total mass and binary fraction.

Our simulated YSCs are expected to dissolve in the galactic disc in $\sim 100$ Myr, 
that is much shorter than the coalescence timescale of all BH-BH binaries and of some NS-NS binaries. 
The DCOBs that form within the simulated YSCs are ejected in the field 
(due to three-body interactions or because of the disruption of the parent YSC). 
Once in the field, the DCOBs will not undergo more dynamical interactions and 
will continue their evolution in isolation, until they merge. Thus, the mergers 
of (most) our simulated DCOBs are expected to take place in the field. Accounting for the fact that 
most DCOBs form in YSCs and evolve through dynamical interactions is a crucial 
step towards obtaining a realistic description of the demographics 
 of DCOBs, in light of the forthcoming Advanced LIGO and VIRGO scientific runs.




\section*{Acknowledgments}

We thank the anonymous reviewer for the careful reading 
and the constructive comments which helped us 
to improve the manuscript. We also thank Alessandra Mastrobuono Battisti and Roberto Soria for useful discussions. We made use of the public software package {\sc starlab} (version 4.4.4) and of the {\sc sapporo} library (\citealt{gaburov09}) to run {\sc starlab} on graphics processing units (GPUs). We acknowledge all the developers of {\sc starlab}, and especially its primary authors: Piet Hut, Steve McMillan, Jun Makino, and Simon Portegies Zwart. We thank the authors of {\sc sapporo}, and in particular E. Gaburov, S. Harfst and S.  Portegies Zwart.
 BMZ, MM and MB acknowledge financial support from the Italian Ministry of Education,
University and Research (MIUR) through grant FIRB 2012 RBFR12PM1F. BMZ was supported by a PhD fellowship from Padova University through the
Strategic Research Project AACSE (Algorithms and Architectures for Computational
Science and Engineering). 
MM acknowledges financial support from INAF through grant PRIN-2011-1 and from CONACyT 
through grant 169554. We acknowledge the CINECA Award N. HP10B3BJEW, HP10CLI3BX,
HP10CXB7O8, HP10C894X7, HP10CGUBV0, HP10CP6XSO and HP10C3ANJY for
the availability of high performance computing resources and support.\\

\appendix

\section{Stable versus unstable DCOBs}
As we mentioned in Section~\ref{sec:DBHpopulation}, in our paper a binary system is defined as a bound pair, i.e. the most general possible definition. On the other hand, it is reasonable to expect that a portion of these binaries are extremely loose systems, which remain bound only for one (or few) time-steps (see the discussion in Section~\ref{sec:OrbitalProperties}). In this Appendix, we discuss how our results are influenced by our definition of binary systems. In particular, we will compare the main properties of stable and unstable DCOBs.

{\sc starlab} defines as stable binaries those bound pairs with periastron distance $r_{\rm p}\le{}2.5\,{}R_{\rm close}$ (see \citealt{pz2001}), where $R_{\rm close}$ is defined as
\begin{equation}\label{eq:stabilityCriterion}
 R_{\rm close} = r_{\rm vir}\frac{m_1+m_2}{2\,{}M_{\rm tot}}.
\end{equation}
 Then, unstable binaries are binaries with periastron  $r_{\rm p}>2.5\,{}R_{\rm close}$. In the following, we consider stable and unstable binaries separately.

\begin{figure}
\includegraphics[width = \columnwidth]{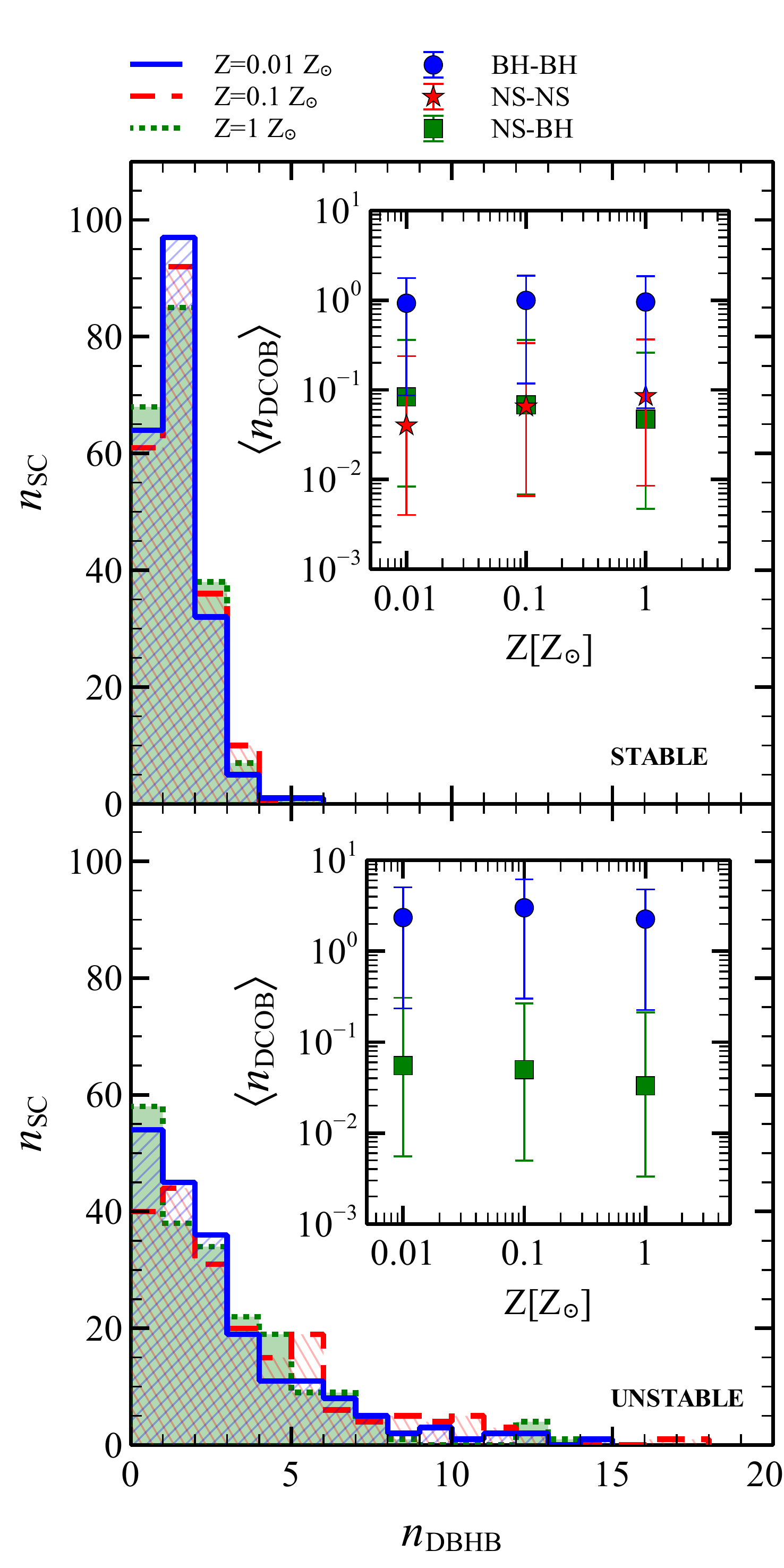}
\caption{The same as Fig.~\protect\ref{fig:meanDCOBperZ_and_distro}, but we distinguish between stable (top) and unstable (bottom) binaries.}
\label{fig:appendixDistro} 
\end{figure}


\subsection{DCOB population}

\begin{figure}
\includegraphics[width = \columnwidth]{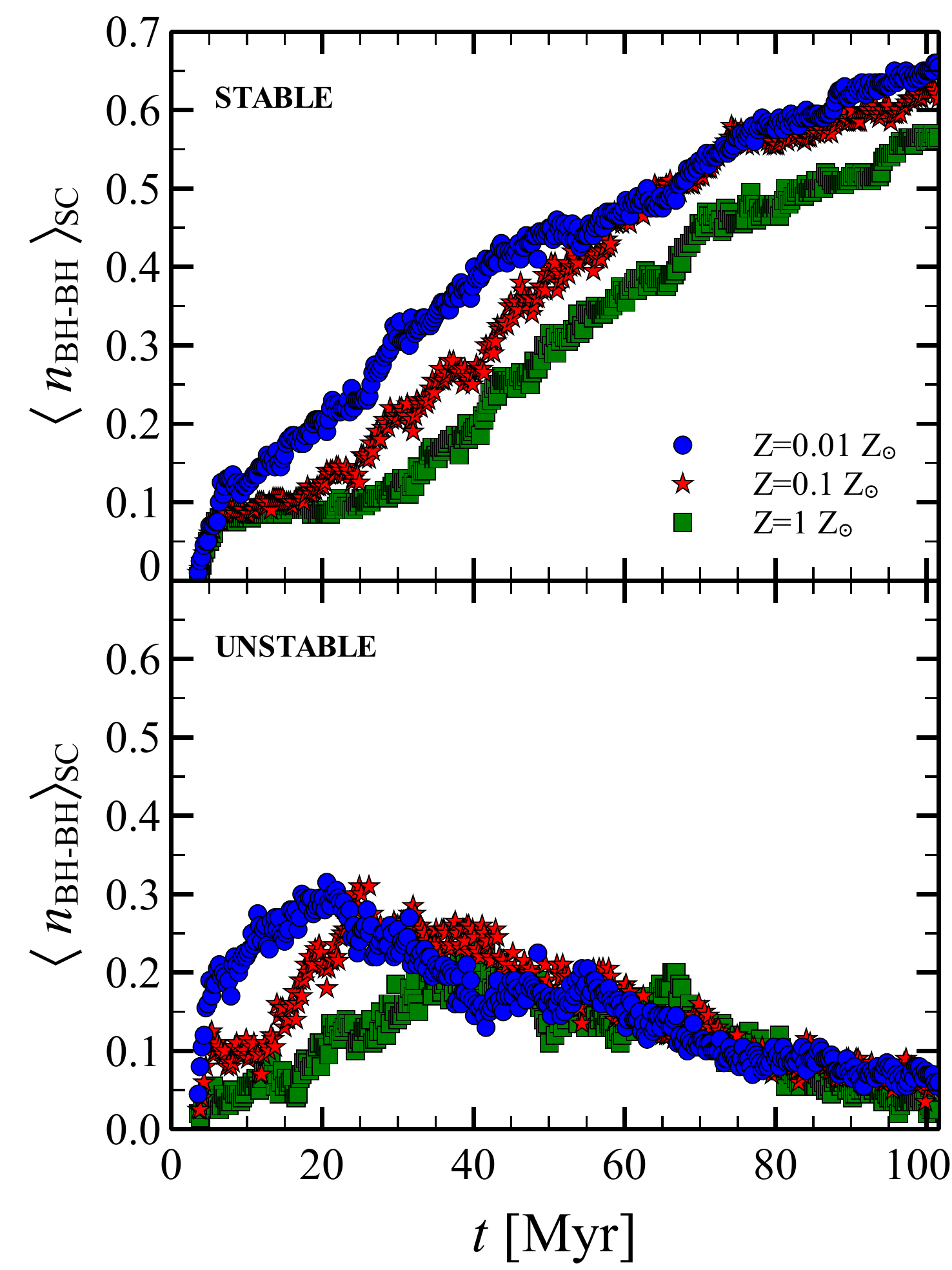}
\caption{The same as Fig.~\protect\ref{fig:vCountInTime}, but we distinguish between stable (top) and unstable (bottom) binaries.}
\label{fig:appendixTime} 
\end{figure}
Fig.~\ref{fig:appendixDistro} is the same as Fig.~\ref{fig:meanDCOBperZ_and_distro}, but it has been derived considering stable and unstable binaries separately (in the top and bottom panel, respectively). The inset of Fig.~\ref{fig:appendixDistro} shows the average number of BH-BH, NS-NS and 
NS-BH per YSC as a function of the metallicity. It is remarkable that  
BH-BH binaries are at least ten times more numerous than  NS-NS and NS-BH binaries, when considering both the stable binary sample and the unstable binary sample. This shows that dynamics has a strong impact on the population of DCOBs, regardless of the distinction between stable and unstable binaries. It is also worth noting that we have found no unstable NS-NS binaries. 
This confirms that only hard (stable) NS-NS binaries can survive (without being disrupted) the two SN explosions of the two progenitors and the dynamical evolution of the binary.

The main panel of Fig. \ref{fig:appendixDistro} shows the distribution of BH-BH binaries per YSC (integrated over the simulation time). Here, the difference between stable and unstable binaries is quite marked: a single YSC can host up to $\sim{}18$ unstable binaries, but only up to $\sim{}6$ stable binaries. 

Fig. \ref{fig:appendixTime} compares the average number of BH-BH binaries per YSC as a function of time for stable (top) and unstable (bottom) binaries. 
It is worth noting that unstable binaries peak at $10\,{}{\rm Myr}<t<40\,{}{\rm Myr}$, i.e. immediately after the core collapse: it is reasonable to expect that the formation of loose binaries is triggered by the increase of the central density due to the core collapse phase (see \citealt{mapelli2013}). In contrast, the number of stable binaries steadily increases with time (because they tend to survive for a longer time, after their formation). The differences among metallicities that we discussed in Section~\ref{sec:DBHpopulation} still hold, when considering stable and unstable binaries separately. 

\subsection{Orbital properties and coalescence timescale}

\begin{figure}
\includegraphics[width = \columnwidth]{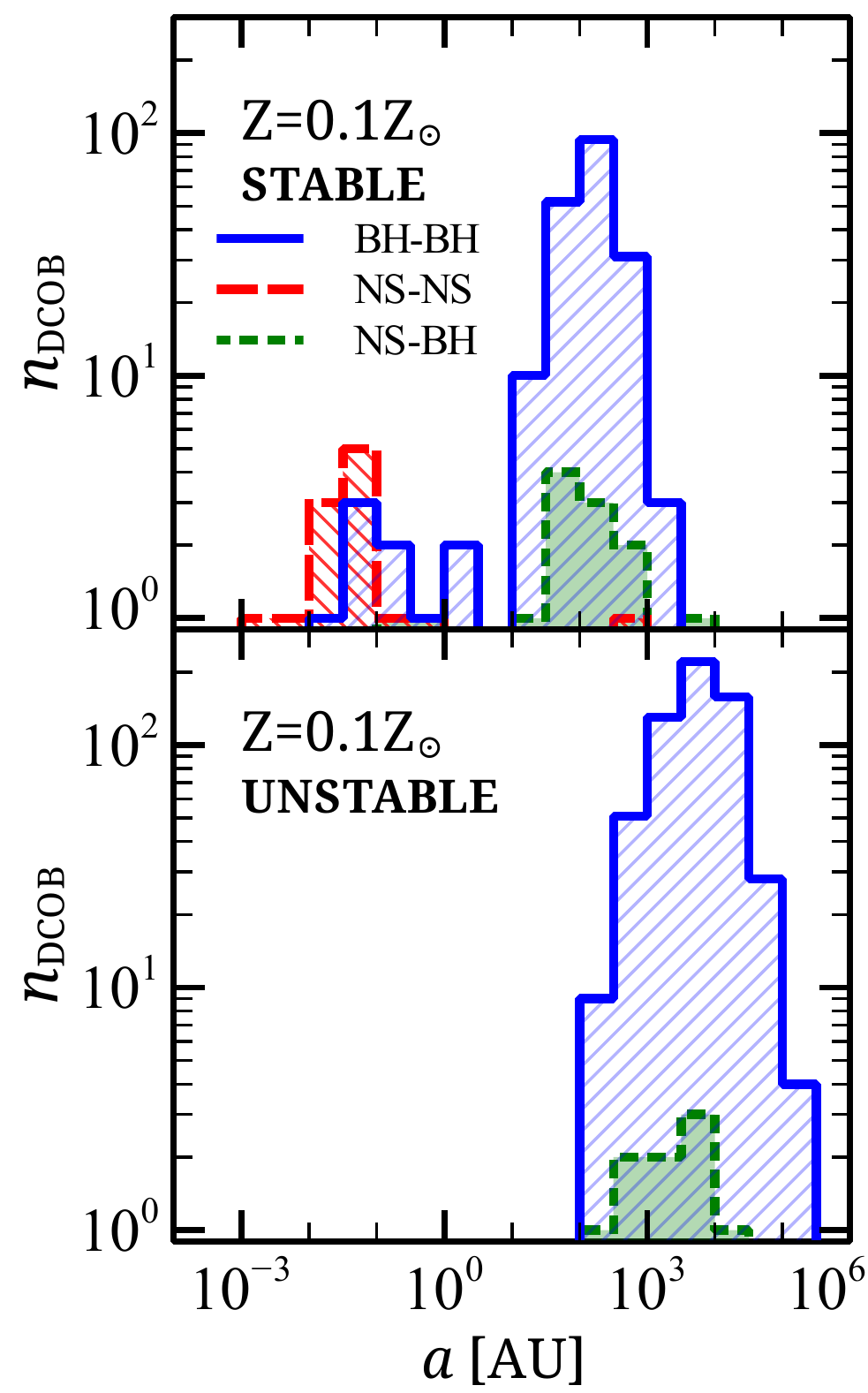}
\caption{Distribution of semi-major axes $a$ for the stable (top) and unstable (bottom) DCOBs at $Z=0.1$ Z$_\odot$. Lines and colours are the same as in Fig.~\ref{fig:vOrbPropHist}.}
\label{fig:appendixSma} 
\end{figure}
Fig. \ref{fig:appendixSma} shows the distribution of semi-major axes of BH-BH, NS-NS  and NS-BH binaries at $Z=0.1$ Z$_\odot{}$, distinguishing between stable (top) and unstable (bottom) binaries. 
As it is reasonable to expect, most unstable (stable) binaries 
have semi-major axes $>10^3$ AU ($<10^3$ AU). However, there are also some unstable binaries with $a$ smaller than that of stable binaries. The reason is that the stability criterion depends not only on the separation of the two objects, but also on their mass (in this sense, it is a hardness criterion) and eccentricity.

In the bottom panel of Fig. \ref{fig:appendixSma}, the most loose unstable binaries have semi-major axes as large as $10^6$ AU, that is $\sim 5$ pc (similar to the initial YSC tidal radius), with periods comparable to the initial central two-body relaxation time ($\sim $ 10 Myr, see also Fig. \ref{fig:vOrbPropHist}). These extremely loose bound pairs are very short-lived: it is reasonable to expect that they would completely disappear, if a galactic tidal field would be included in our simulations. On the other hand, these highly unstable systems are completely negligible from the point of view of GW sources.

\begin{figure}
\includegraphics[width = \columnwidth]{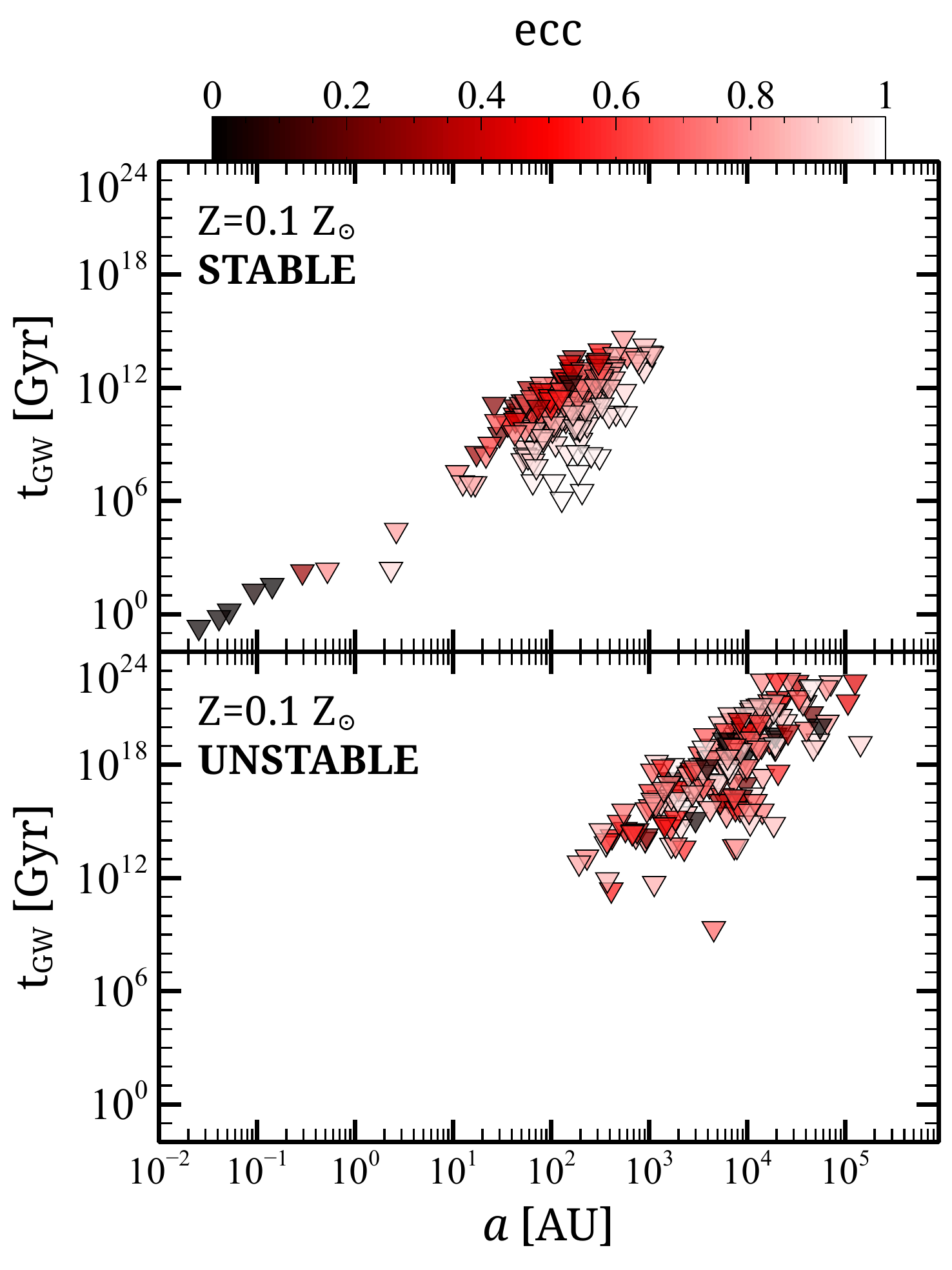}
\caption{Coalescence timescale as a function of the semi-major axis for stable (top) 
and unstable (bottom) BH-BH binaries at $Z=0.1$ Z$_\odot$. Symbols and colours are the same as in Fig.~\ref{fig:vPetersTimesMap}.}
\label{fig:appendixGW} 
\end{figure}

Fig. \ref{fig:appendixGW} confirms that unstable DCOBs are completely negligible from the point of view of GW emission: their coalescence timescale is by orders of magnitude longer than the Hubble time. Thus, it is sufficient to consider stable binaries alone, when we are interested in possible GW sources. 

Finally, in this Section we have considered only YSCs with $Z=0.1$ Z$_\odot$ as an example.  The same conclusions can be drawn for the other metallicities.





\end{document}